\documentclass[aip,reprint,jcp,groupedaddress]{revtex4-1}
\pdfoutput=1

\usepackage[]{natbib}

        \usepackage[usenames,dvipsnames]{color}
        \usepackage[pdftex]{graphicx}
        \usepackage{epstopdf}
        \DeclareGraphicsExtensions{.pdf, .eps, .png}

\begin{document}
%
\title{Coarse-grained model of adsorption of blood plasma proteins onto nanoparticles}
\author{Hender Lopez}
\email[Electronic mail: ]{hender.lopezsilva@ucd.ie}
\author{Vladimir Lobaskin}
\email[Electronic mail: ]{vladimir.lobaskin@ucd.ie}
\affiliation{School of Physics, Complex and Adaptive Systems Lab, University College Dublin, Belfield, Dublin 4, Ireland}

\begin{abstract}
{We present a coarse-grained model for evaluation of interactions of globular proteins with nanoparticles. The
protein molecules are represented by one bead per aminoacid and the nanoparticle by a homogeneous sphere that interacts
with the aminoacids via a central force that depends on the nanoparticle size. The proposed methodology is used to predict the 
adsorption energies for six common human blood plasma proteins on hydrophobic charged or neutral nanoparticles of different sizes as well
as the preferred orientation of the molecules upon adsorption. Our approach allows one to rank the proteins by their binding affinity to 
the nanoparticle, which can be used for predicting the composition of the NP-protein corona. The predicted ranking is in good agreement 
with known experimental data for protein adsorption on surfaces. }
\end{abstract}
\maketitle

\section{Introduction} \label{sec::intro}

When uncoated nanoparticles (NP) enter a living organism, they are first exposed to biological fluids,
which results in a formation of stable or transient NP-biomolecule complexes. For large NPs, the biomolecular coating is
referred to as (protein) corona.     
It has been shown that composition and structure of the corona determines the biological reactivity and toxicity of the
NPs~\cite{MASD2012,1-72,1-72a,1-72b,1-72c} as well as the NP systemic transport including NP uptake into cells. The content of the corona
can be directly linked to the toxic effects and used to predict the toxicity of engineered nanomaterials.\cite{Kamath2015}
In addition to the study of possible hazards, the interest to nanobio interactions is driven by promising
applications of NPs in food, cosmetics, and medicine.\cite{Salata2004,Rahman2012,Wang2014} A quantitative model of NP corona formation 
can facilitate designing of new nanomaterials with specific functions.

The composition of the protein corona and protein adsorption kinetics have been studied extensively by a broad
range of experimental techniques such as fluorescence correlation spectroscopy (FCS),\cite{Roeckeretaal2009}
differential centrifugal sedimentation (DCS) combined with imaging techniques,\cite{Keatal2015}
quantitative liquid chromatography mass spectrometry (LC-MS)~\cite{Silva2006,Ritz2015} and
dynamic light scattering (DLS) combined with isothermal titration calorimetry (ITC)~\cite{Winzen2015}
(for a recent review see Ref. \cite{Pino2014}). Despite the valuable information that 
experimental techniques have provided, there is still much controversy and gaps in the physical picture of protein
adsorption on NPs: disagreement on whether the adsorption is reversible, whether proteins
change conformation and preserve their functionality when complexed with certain particle type, whether the corona survives the NP uptake into the cell, etc.
Undoubtedly, computer simulations can assist experimental data and reveal molecular scale information
required to understand the corona formation process.

Full atomistic simulation of protein on surfaces have already proved useful to advance the understanding of 
molecular interactions that determine the binding of proteins to inorganic nanoparticles.\cite{BKCW2012,DRCC2013,KGN2013,Tavanti2015}
The atomistic simulations are however limited to systems composed of one or few proteins and give
information well below the time scales relevant for the formation of the protein corona.
A solution to overcome this restriction is to use coarse-grained (CG) models that reduce
the number of interaction sites used in the simulation but keep the required molecular information about
the proteins and the NPs. Some CG models to study the kinetics of the protein corona formation have already
been proposed (see Vilaseca {\it et al.},\cite{VDF2013} Bellion {\it et al.},\cite{Bellion2008}
Oberle {\it et al.}~\cite{Oberle2015}, as well as the section on computer simulations of the review 
by Rabe {\it et al.}~\cite{Rabeetal2011}), but most of these works use rather simplistic presentation of the proteins and lack
molecular detail that could be essential for the adsorption kinetics.

In this work, we develop a CG model that allows us to calculate the adsorption energies
of arbitrary globular proteins onto hydrophobic NPs of arbitrary size. The model is built starting from the molecular
structure of the proteins, and the size of the NP is explicitly included in the model. In Section~\ref{sec::model}
we give a detailed description of our model and describe the parametrization process. In Section~\ref{sec::results}
we show the numerical results on adsorption of six most abundant human blood plasma proteins on NPs
of different radii and charge. Finally, in Section~\ref{sec::conclusions} we present the conclusions.

\section{Model} \label{sec::model}

\subsection{Coarse-grained protein model}\label{sec::protein}

\begin{figure}[tbh]
\centering
\includegraphics[width=\hsize]{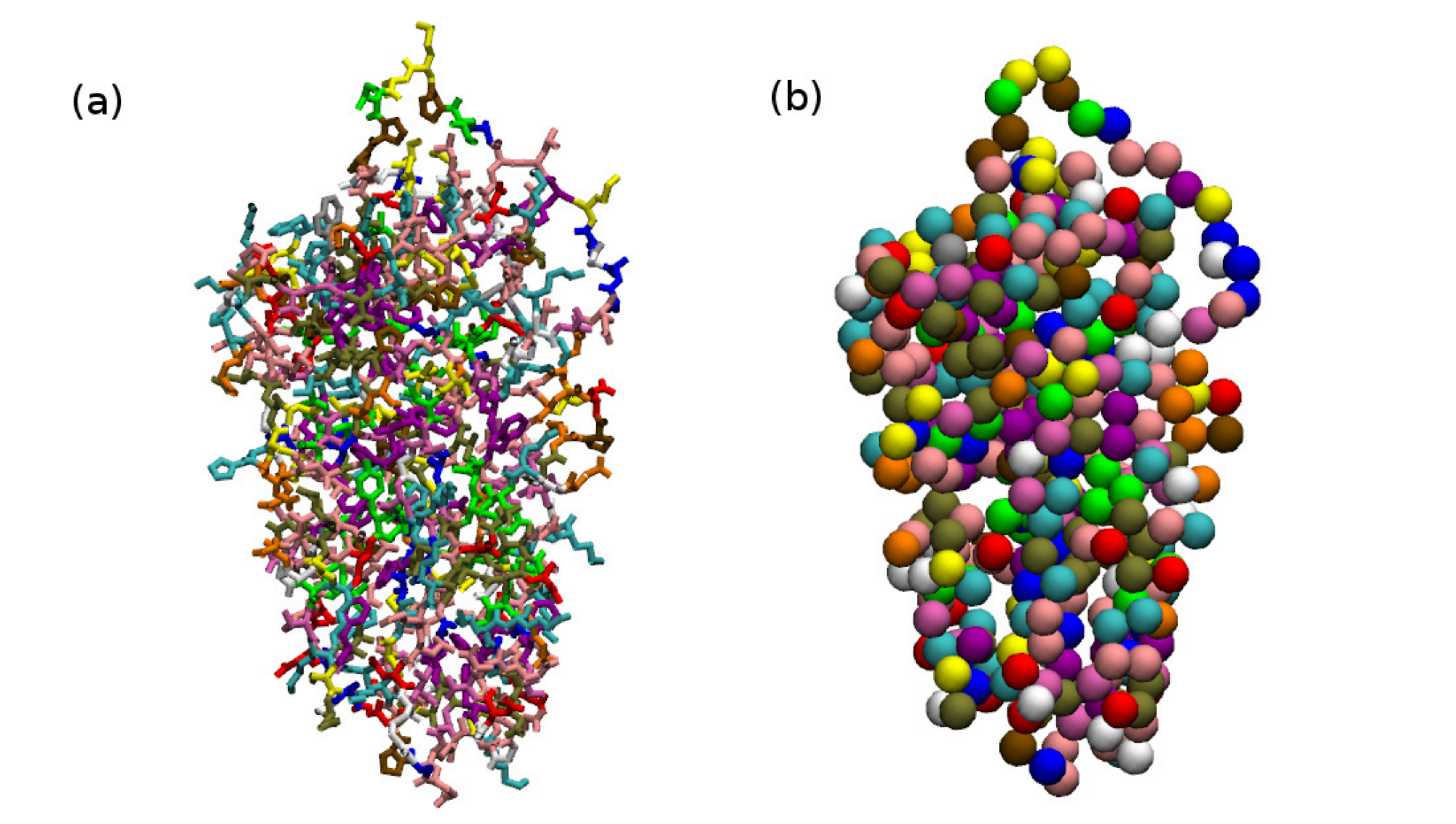}
\caption{Two representations of the $\alpha_1$-antitrypsin (A1A) molecule. (a) A full atoms representation taken from
the PDB file ID: 3NE4 and (b) the one-bead-per-aminoacid model proposed in this work.}
\label{fig::CGfull}
\end{figure}

Our main aim is to design a CG model that would reflect NP-protein and protein-protein interactions 
and could be scaled up to simulate multiple biomolecules in contact with a NP on relatively long
times. Much work has been done recently on systematic coarse-graining of the proteins (for reviews see~\cite{Toz2005,Tak2012,Noid2013}).
Based on the previous experience, we propose a single-bead-per-aminoacid model and consider each protein molecule as a rigid body. 
This model preserves the shapes and sizes of the proteins (and therefore, their mobilities and excluded volume effects) as well as the 
surface charge distribution, so we hope to be able to address their competitive adsorption on the NP surface.
We use crystal structures of the proteins as obtained from the Protein Data Bank (PDB) and place one bead for each
aminoacid at the position of the corresponding $\alpha$-carbon atom. Fig.~\ref{fig::CGfull} shows an example of
coarse-graining of protein $\alpha_1$-antitrypsin (A1A) as taken from the PDB file ID: 3NE4 and our one-bead-per-aminoacid
CG model. For the NP, we will consider here only spherical homogeneous objects, so that a single bead presentation is
sufficient.

\subsection{Nanoparticle-aminoacid interactions}\label{sec::potentials}

We assume pairwise additivity of all interactions and present the net NP-protein interaction energy ($U$) as a sum of the individual interactions of the amnoacids that
compose the protein with the NP material. Furthermore, $U$ is not only a function of distance from the surface of the NP to the center of
mass (COM) of the protein, $d_{\mathrm{COM}}$, but also depends on the protein orientation, which
is characterized by two Euler angles $\phi$ and $\theta$ (defined in the next section).
In our model, $U(d_{\mathrm{COM}},\phi,\theta)$ includes two contributions:
\begin{equation}\label{eq::U}
U(d_{\mathrm{COM}},\phi,\theta) = \sum_{i=1}^{N} \left( U_i^{\mathrm{VdW}} + U_i^{\mathrm{el}}\right),
\end{equation}
where $N$ is the number of aminoacids in the protein, $U_i^{\mathrm{VdW}}$ is the van der Waals interaction
of aminoacid $i$ with the surface and $U_i^{\mathrm{el}}$ is the electrostatics interaction of the aminoacid $i$ with the surface.

As we are interested in studying the effect of the size of the NP on the adsorption energy, we present the
van der Waals interaction potential in a form that explicitly includes the radius of the NP as a parameter, following the
well known Hamaker procedure.\cite{Hamaker1937}
We start from the residue-residue interaction potential proposed by Bereau and Deserno,\cite{BerDes2009} which
is based on a modified 12-6 Lennard-Jones potential. In this model, each aminoacid residue is characterized by the hydrophobicity index ($\epsilon_i$)
and we additionally assume that any surface segment of the nanomaterial can be modelled in the same way as the aminoacid ($\epsilon_s$).
With these assumptions, the interaction between the aminoacid $i$ and a bead of the nanomaterial $s$ being at
a distance $r$ from each other is given by:
\begin{widetext}
\begin{equation}\label{eq::LJRS}
U_{s,i}(r) =
\left\{
        \begin{array}{ll}
                4\epsilon_{en}\left[\left(\frac{\sigma_{s,i}}{r}\right)^{12}-\left(\frac{\sigma_{s,i}}{r}\right)^{6}\right]
                +\epsilon_{en}(1-\epsilon_{s,i}), & r < r_{c,i},\\
               4\epsilon_{en}\epsilon_{s,i}\left[\left(\frac{\sigma_{s,i}}{r}\right)^{12}-\left(\frac{\sigma_{s,i}}{r}\right)^{6}\right],
                & r_{c,i}\leq r \leq r_{\mathrm{cut}},\\
                0, & r >  r_{\mathrm{cut},}
        \end{array}
\right.
\end{equation}
\end{widetext}
where $\epsilon_{en}$ is a parameter that scales the interaction energy, $\epsilon_{s,i}$ is the combined hydrophobicity index of residue $i$ and the nanomaterial according to the usual Lorentz-Berthelot mixing
rules and is given by $\epsilon_{s,i}=\sqrt{\epsilon_s \epsilon_i}$, $\sigma_{s,i}$ is the average van der Waals radius of residue $i$ and the NP bead, $\sigma_{s,i}=(\sigma_s + \sigma_i)/2$, and $r_{c,i}$ is the position of the minimum of the pair potential.
We follow the same methodology as Ref.~\cite{BerDes2009} to define the hydrophobicity index, which is based on the widely
used residue-residue interaction energies proposed
by Miyazawa and Jernigan,\cite{MiyJer1996} but instead of having a $20 \times 20$ interaction matrix
this is reduced to a table of hydrophobicities, one for each aminoacid
(see Table~II in \cite{BerDes2009}). A hydrophobicity index 0 is assigned to the most hydrophilic
residue (LYS) and an index 1 to the most hydrophobic one (LEU).
At this point, we should stress that any other hydrophobicity scale can also be used,
with the only condition that it has to be transformed in such a way that the indexes lay
between 0 and 1. 

In the above expression, we consider only a small volume element of nanomaterial, similar to an aminoacid in scale,  but to further coarse-grain the interaction we
integrate the energy over a semi-infinite volume of the material (with a flat surface) and then over a spherical NP.
For a flat surface, the interaction potential can be expressed in terms of $d$, the distance between the residue center of mass and the closest element of the surface.
An integration of the 12-6 potential defined in Eq.~(\ref{eq::LJRS}) over a semi-space gives:
\begin{widetext}
\begin{equation}\label{eq::LJRS1}
U_{i}^{\mathrm{vdW}}(d) =
\left\{
        \begin{array}{ll}
                 \epsilon_{es} \rho \sigma_{s,i}^3 \left[\left(\frac{\sigma_{s,i}}{d}\right)^{9}-\frac{15}{2}\left(\frac{\sigma_{s,i}}{d}\right)^{3}
                +\left(\frac{125}{2}\right)^\frac{1}{2}(1-\epsilon_{s,i})\right ], & d < d_{c,i},\\
                 \epsilon_{es}  \epsilon_{s,i} \rho  \sigma_{s,i}^3 \left[\left(\frac{\sigma_{s,i}}{d}\right)^{9}-\frac{15}{2}\left(\frac{\sigma_{s,i}}{d}\right)^{3}\right], & d_{c,i}\leq d \leq d_{\mathrm{cut}},\\
                0, & d >  d_{\mathrm{cut},}
        \end{array}
\right.
\end{equation}
\end{widetext}
where $\epsilon_{es} = \frac{4 \pi }{45}\epsilon_{en}$, $\rho$ is the number density of beads in the nanomaterial, $d$ is the distance from the residue $i$ to the surface,
$d_{c,i} = (2/5)^{1/6}\sigma_{s,i}$. Although the density $\rho$ seems to be an important parameter scaling the interaction, it is not an independent quantity and
therefore is not crucial for our method. From fitting the adsorption energy to experimental or MD simulation data, we can find the composite quantity $\epsilon_{es} \rho$ (energy density),
which is sufficient for further calculations. For a nanoparticle of radius $R$, a similar integration over the particle volume gives:
\begin{widetext}
\begin{equation}\label{eq::LJRS2}
U_{i}^{\mathrm{vdW}}(r) =
\left\{
        \begin{array}{ll}
                4 \epsilon_{es} \rho \sigma_{s,i}^3  \left[\frac{\left ( 15 r^6 R^3+ 63 r^4 R^5 + 45 r^2 R^7 + 5 R^9 \right) \sigma^{9}_{s,i}} {\left(r^2 - R^2\right)^9}
                - \frac{15 R^3 \sigma^{6}_{s,i}} {\left (r^2 - R^2 \right )^3} \right]
                - U^{\mathrm{vdW}}_c (1-\epsilon_{s,i}), & r < r_{c,i},\\
                4 \epsilon_{es}\epsilon_{s,i} \rho \sigma_{s,i}^3 \left[\frac{\left ( 15 r^6 R^3+ 63 r^4 R^5 + 45 r^2 R^7 + 5 R^9 \right) \sigma^{9}_{s,i}} {\left(r^2 - R^2\right)^9}
                - \frac{15 R^3 \sigma^{6}_{s,i}} {\left (r^2 - R^2 \right )^3} \right], & r_{c,i}\leq r \leq r_{\mathrm{cut}},\\
                0, & r >  r_{\mathrm{cut},}
        \end{array}
\right.
\end{equation}
\end{widetext}
where $r$ is the distance from aminoacid $i$ to the center of the nanoparticle. The distance $r_{c,i}$ corresponds to the minimum of the potential and $U^{\mathrm{vdW}}_c$ is the value of the function $U_{s,i}^{\mathrm{vdW}}(r_{c,i})$ as defined in the range $r_{c,i}\leq r \leq r_{\mathrm{cut}}$. We do not show the general expression for the position of the minimum as it is too bulky. The minimum is located at
$r_{c,i} - R \approx(2/5)^{1/6}\sigma_{s,i}$ at $R \gg \sigma_{s,i}$ and is moved to shorter distances at smaller $R$. The variation, however, is not very large, at $R \to \infty$, $r_{c,i} - R \approx 0.858374 \sigma_{s,i}$, at $R = 200 \sigma_{s,i}$ it is  $0.858375 \sigma_{s,i}$, at $R = 20 \sigma_{s,i}$ it is $0.858469 \sigma_{s,i}$, and at $R = 2 \sigma_{s,i}$ it is $0.865242 \sigma_{s,i}$.

\begin{figure}[tbh]
\centering
\includegraphics[width=\hsize]{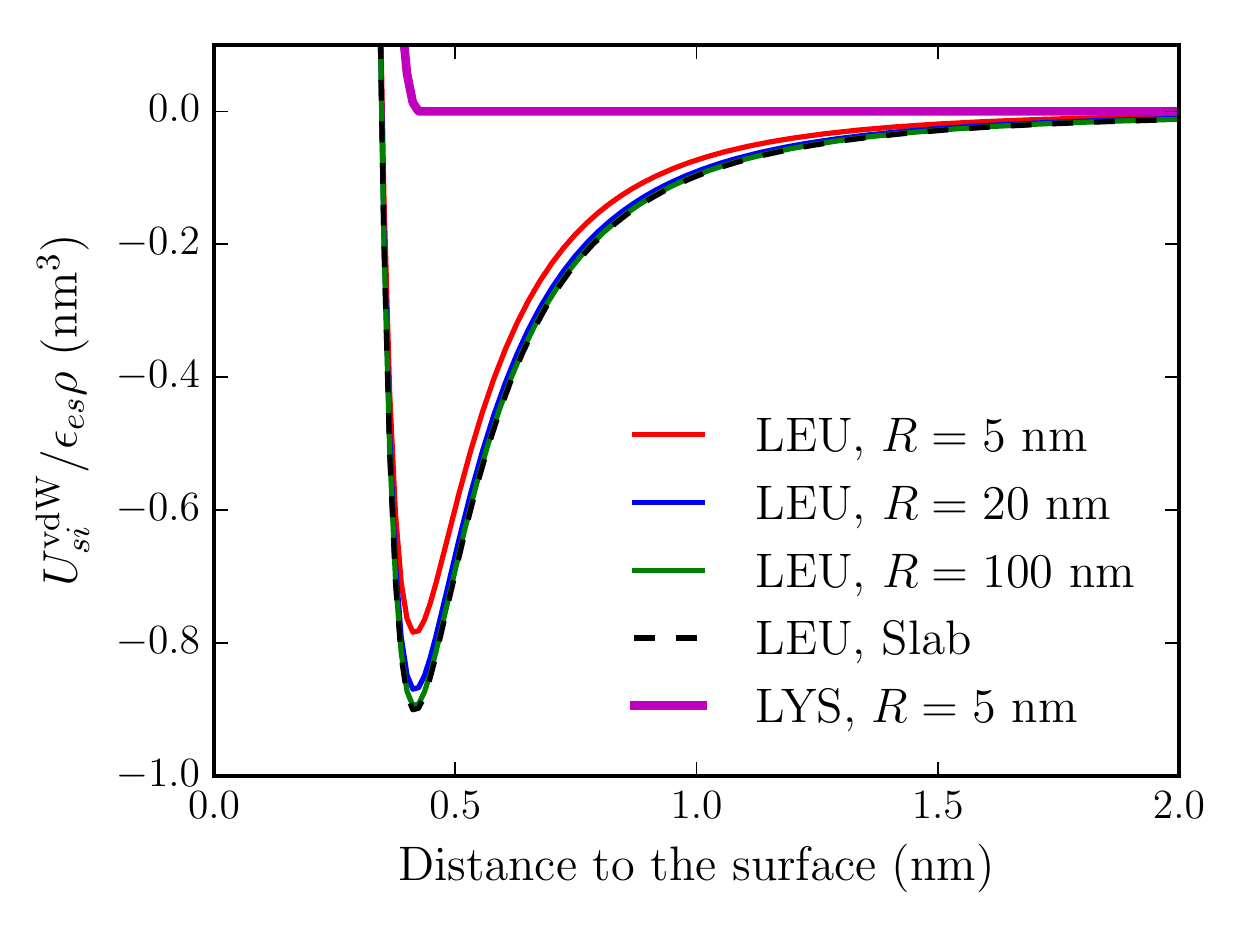}
\caption{Van der Waals interaction potential divided by $\epsilon_{es} \rho$ as a function of the distance of an aminoacid to the adsorbing surface with $\epsilon_s=1$
calculated using Eq.~(\ref{eq::LJRS2}). Residue LEU ($\epsilon_i=1$) for three NP radii ($R=5$~nm, 20~nm and 100~nm)
and a slab are shown. Residue LYS ($\epsilon_i=0$) for a NP with $R=5$~nm.}
\label{fig::potentials}
\end{figure}

Note that the potential proposed (Eqs.~(\ref{eq::LJRS}) and~(\ref{eq::LJRS2})) will only give a repulsive interaction between a
highly hydrophilic surface and any aminoacid residue (\textit{i.e.} defining $\epsilon_s=0$, gives $\epsilon_{s,i}=0$ for all residues).
On the other hand, assigning a non-zero value for $\epsilon_s$ will only change the magnitude of
the interaction between the aminoacid and the surface but not the shape of the potential.
In this way, the proposed potential is limited to only hydrophobic surfaces and cannot reproduce
the attraction  between the beads across the hydration layer.
Because of this limitation, we set the value of $\epsilon_s=1$ for all simulations. As an illustration,
Fig.~\ref{fig::potentials} shows the proposed van der Waals potential for LEU ($\epsilon_i=1$) and LYS ($\epsilon_s=0$).
Alternatively, a potential that includes hydration effects, such as a 12-10-6 Lennard-Jones potential for residue-residue
interactions~\cite{KTMH2008,KimHum2008} or the modified version proposed by Wei and Knotts~\cite{WeiKno2013}
to model residue-surface interactions can be used to generate a more general interaction potential.
The main drawback of the use of these more refined formulas for the potential is that the parametrization
is more challenging and the applicability of a set of parameters can be very narrow.

The electrostatic interactions in Eq.~(\ref{eq::U}) are modeled by placing point charges on the NP surface.
This charges interact with the charged groups of the protein via a Debye-H\"uckel potential.
The electrostatic interaction energy between an aminoacid $i$ and all the charges on the surface is
given by:
\begin{equation}\label{eq::DHpotential}
U^{el}_{i} =  \sum_{j=1}^{N_e} \lambda_B k_B T q_i q_j \frac{\exp(-r_{ij}/\lambda_D)}{r_{ij}},
\end{equation}
where $r_{ij}$ is the distance between the residue $i$ and the point charge $j$ on the surface,
$\lambda_B = e^2/ \left(4\pi\varepsilon_0 \varepsilon_r k_B T\right)$ is the Bjerrum length, $k_B$ is the Boltzmann constant,
$T$ the temperature, $\varepsilon_0$ the dielectric permittivity of vacuum, $\varepsilon_r$ the relative  dielectric permittivity of water,
$q_i$ the charge of residue $i$, $q_j$ the charge of the point charge $j$ on the surface, $N_e$ the total
number of point charges on the surface and $\lambda_D$ is the Debye length
(defined through $\lambda_D^{-2} = 8 \pi \eta \lambda_B c_0$, with $c_0$ is the background
electrolyte concentration). In practice, the points charges are evenly distributed on the spherical surface of the NP using
a Golden Section spiral algorithm and all points will have the same charge $q_j$
given by $q_j = 4 \pi \sigma R^2/N_e$, where $\sigma$ is the surface charge density of the NP
and $R$ is the radius of the NP.

\begin{table*}
\begin{tabular}{c | c c c c c c c c c c}
\hline
\hline
Residue & LYS & GYU & ASP & ASN & SER & ARG & GLU & PRO & THR & GLY\\
$\epsilon_i$~$(\mathcal{E})$ & 0.00 & 0.05 & 0.06 & 0.10 & 0.11 & 0.13 & 0.13 & 0.14 & 0.16 & 0.17\\
$\sigma_i$~(nm) & 0.64 & 0.59 & 0.56 & 0.57 & 0.52 & 0.66 & 0.60 & 0.56 & 0.56 & 0.45\\
\hline
\hline
Residue & HIS & ALA & TYR & CYS & TRP & VAL & MET & ILE & PHE & LEU\\
$\epsilon_i$~$(\mathcal{E})$ & 0.25 & 0.26 & 0.49 & 0.54 & 0.64 & 0.65 & 0.67 & 0.84 & 0.97 & 1.00\\
$\sigma_i$~(nm) & 0.61 & 0.50 & 0.65 & 0.55 & 0.68 & 0.59 & 0.62 & 0.62 & 0.64 & 0.62\\
\hline
\hline
\end{tabular}\caption{Normalized hydrophobicities $\epsilon_i$
(taken from Table~II in \cite{BerDes2009} and $\sigma_i$ for each amino acid.
The most hydrophilic residue has a $\epsilon_i$ of 0, while the most hydrophobic
has a value of 1. }
\label{tab::parameters}
\end{table*}

\subsection{Orientational sampling and the calculation of the adsorption energy}
\label{sec::orientation}

Here, we are not considering conformational changes during the adsorption process and treat the
proteins as rigid bodies. Although the adsorption process might lead to conformational changes,
this usually happens at longer times than the molecule reorientation on the surface.\cite{ARS2005}
Then, the adsorption energies calculated here should provide a reasonable insight into the
kinetics of the NP-protein corona formation although may slightly underestimate the energy.

\begin{figure}[tbh]
\centering
\includegraphics[width=\hsize]{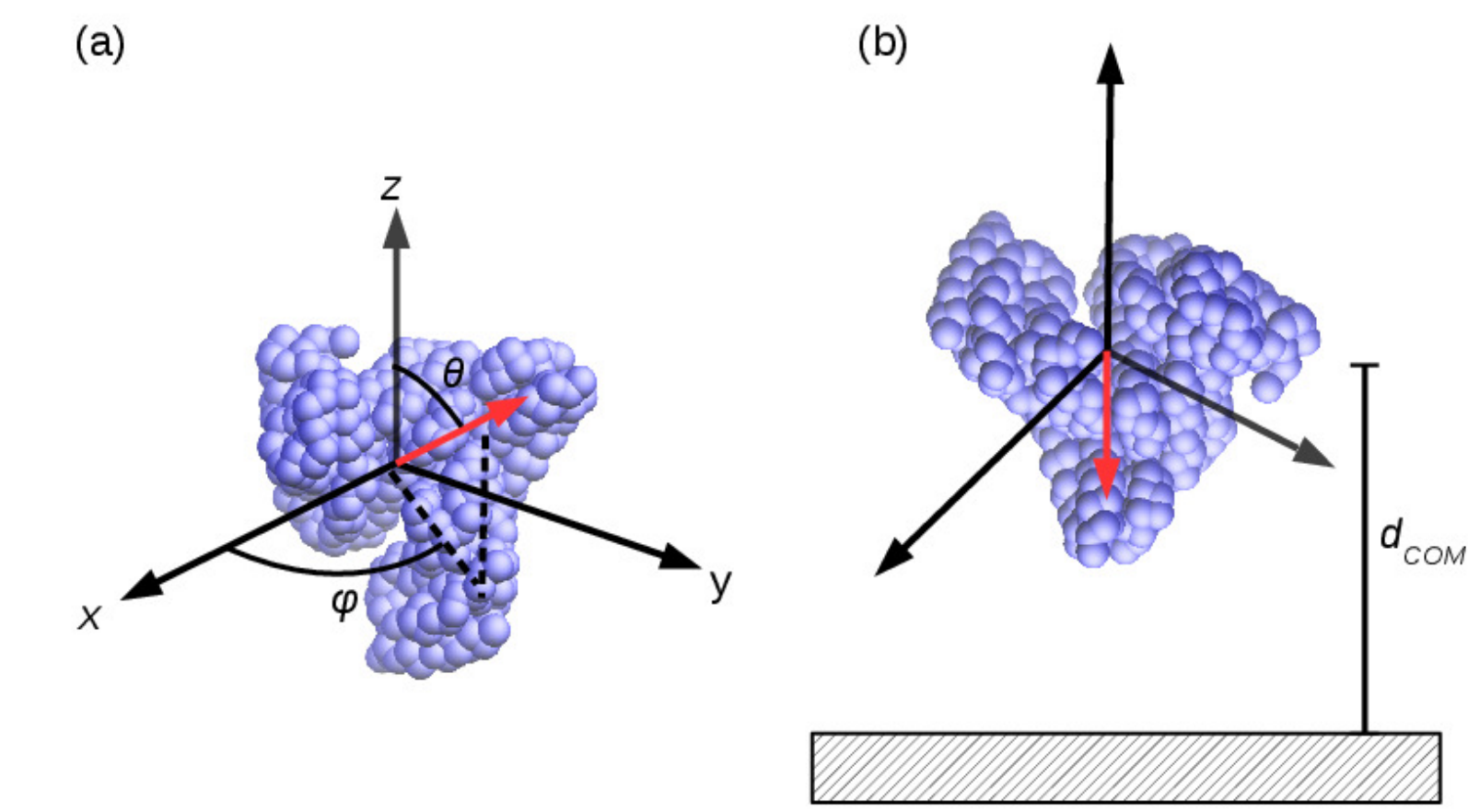}
\caption{Definition of the protein orientation. (a) Any ``atom'' of the protein can be described by
a position vector from the COM, whose orientation is characterized by two angles $\phi$ and $\theta$.
(b) The angles correspond to azimuthal and polar rotations (see Sec.~\ref{sec::orientation})
that would turn the original vector towards the surface (along the negative $z$-axis). The remaining
degree of freedom is the distance of the COM to the surface, $d_{\mathrm{COM}}$.}
\label{fig::confispace}
\end{figure}

To identify the most favorable orientation of adsorbed protein globule (corresponding to the minimum adsorption energy)
we will follow the method suggested by Sun {\it et al.}\cite{SWL2005} Briefly, a configurational space scan is
performed, where a systematic rotation of the protein is used to build a complete adsorption map.
There are three degrees of freedom (DOF) that have to be scanned. Fig.~\ref{fig::confispace} shows that any point 
within the protein molecule can be described by a position vector from the COM of
the protein. This vector is characterized by two angles: $\phi$ and $\theta$ and by
rotating the molecule an angle $-\phi$ about the $z$ direction and then by an angle $-\theta+180\,^{\circ}$
about the $y$ axis will make the position vector point towards the surface (along the negative $z$-axis).
The third DOF is the distance from the COM to the closest point of the surface, $d_{\mathrm{COM}}$.
Here, we sample $\phi$ from 0 to $350^{\circ}$ in steps
of $10^{\circ}$ and $\theta$ from 0 to $170^{\circ}$ in steps of $10^{\circ}$
(note that $\phi=0^{\circ}$ is equivalent
to $\phi=360^{\circ}$, and that $\theta=0^{\circ}$ is equivalent to $\theta=180^{\circ}$).
Instead of obtaining the actual adsorption free energy by calculating the potential of mean force for all orientations,
we only calculate the potential energy $U$ (given by Eq.~(\ref{eq::U})),
which is the sum of all the pairwise interactions between the surface and the aminoacids.
As the net adsorption energies are expected to be well over $k_B T$ and, as the proteins
are assumed to be rigid, neglecting thermal fluctuations is justified.
Note that reference orientations must be chosen to define the angles $\phi$ and $\theta$
$0^{\circ}$. For the simulations reported in this work, the reference orientation
of each protein was the PDB configuration used to build the CG model (more details are given in~\ref{sec::Para}).
For each configuration ($\phi_i$, $\theta_j$), the total potential energy is calculated as
a function of distance of the COM to the surface, $U(d_{\mathrm{COM}},\phi_i,\theta_j)$. Following a similar
approach as in Kokh {\it et al.},\cite{Dariaetal2010} and denoting the reaction coordinate $d_{\mathrm{COM}}=z$,
the mean interaction energy for any particular orientation is given by:
\begin{equation}\label{eq::Esad1}
\begin{array}{l}
E(\phi_i,\theta_j) = -k_B T\\
\times \ln \left[ \frac{1}{a(\phi_i,\theta_j)} \int_0^{a(\phi_i,\theta_j)} \exp\left(\frac{-U(z,\phi_i,\theta_j)}{k_B T}\right) dz \right],
\end{array}
\end{equation}
where $a(\phi_i,\theta_j)$ is the maximum interaction distance from the COM of the protein to the surface for the given orientation.
Then the total mean adsorption energy of the system, $E_{ad}$, can be estimated by averaging over all adsorbed
states with Boltzmann weighting~\cite{SWL2005}:
\begin{equation}\label{eq::Esad}
E_{ad} = \frac{\sum\limits_{i} \sum\limits_{j} P_{ij} E(\phi_i,\theta_j)}{\sum\limits_{i} \sum\limits_{j} P_{ij}},
\end{equation}
where $P_{ij}=\exp[-E(\phi_i,\theta_j)/k_B T]$ is the Boltzmann weighting factor.

\subsection{Details of the simulations, parametrization and validation}
\label{sec::Para}

\begin{table*}
\centering
\begin{tabular}{ c c c c c}
\hline
\hline
Protein & PDB ID & Abbreviation & Weight fraction & Molar mass\\
 &  &  & in plasma, \% & in, kDa\\
\hline
 Human Serum Albumin & 1N5U & HSA & 5.0 & 67 \\
 $\alpha_1$-antitrypsin & 3NE4 & A1A & 0.24 & 51\\
 $\alpha_2$-macroglobulin & 4ACQ & A2M & 0.72 & 725\\
 Fibrinogen & 3GHG & Fib & 0.4 & 340 \\
 Transferrin & 2HAV & Tra & 0.4 & 80 \\
 Immunoglobulin G & 3HR5 & IgG & 1.24 &  150\\
\hline
\hline
\end{tabular}
\caption{Description of proteins used for the model plasma: the abbreviations used
in the text, globule size, and the protein abundance in human blood plasma.}
\label{tab::PDB}
\end{table*}

\begin{figure}[tbh]
\centering
\includegraphics[width=\hsize]{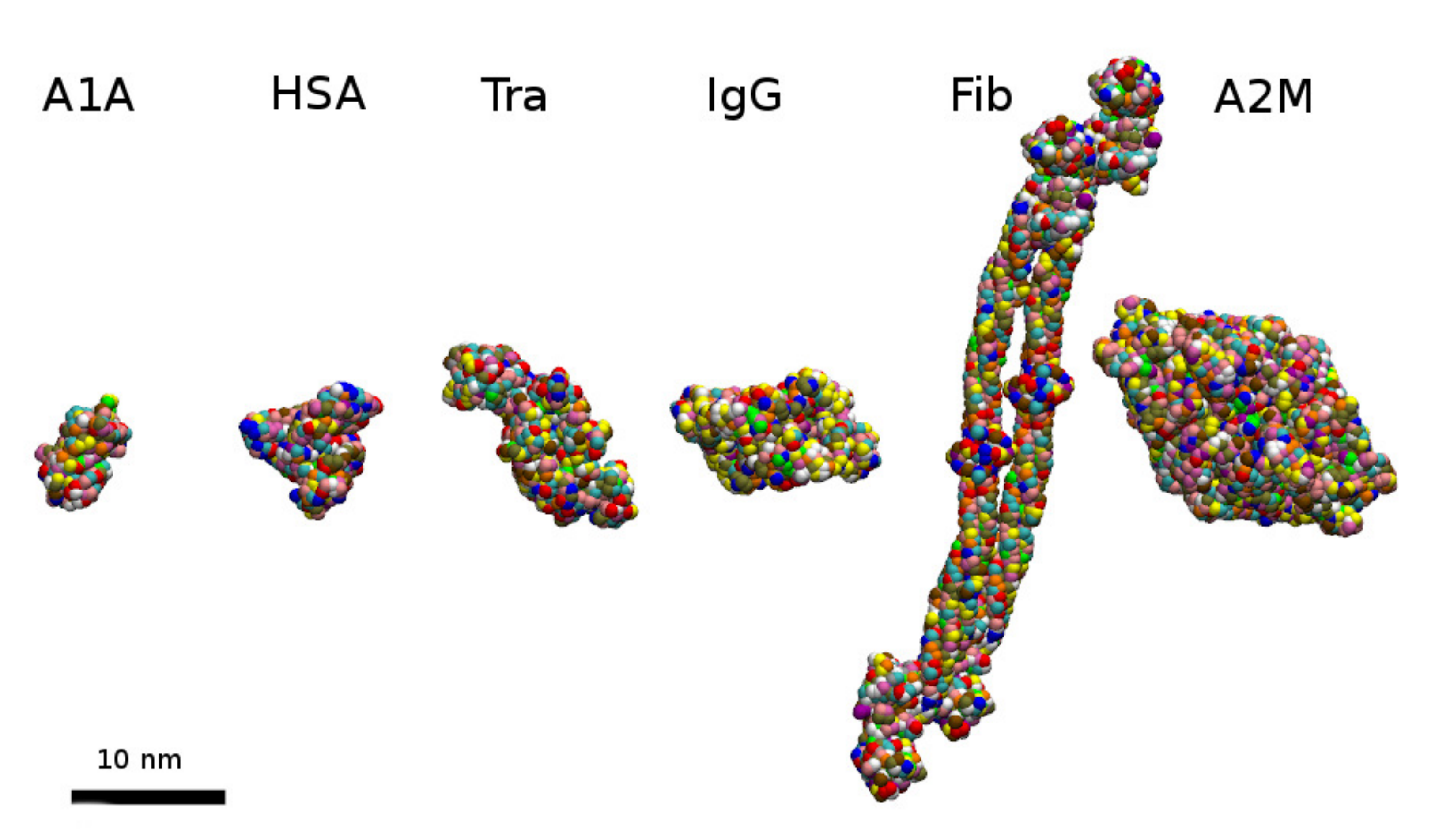}
\caption{CG models of the proteins studied in this work. From left to right: $\alpha_1$-antitrypsin (A1A),
Human Serum Albumin (HSA), Transferrin (Tra), Immunoglobulin G (IgG), Fibrinogen (Fib) and $\alpha_2$-macroglobulin (A2M).
The PDB ID structure from which each CG model was built from are reported in Table~\ref{tab::PDB}.}
\label{fig::proteins}
\end{figure}

Due to complexity of blood plasma, here we will only consider the molecules that are most likely to affect the NP interactions and aggregation and mediate the NP interaction with the cell membranes. The plasma can then be modelled a solution of biomolecules in an implicit solvent with a dielectric constant of water and the Debye length corresponding to physiological ionic strength, van der Waals interactions set to corresponding triplets NP-protein-water, or protein-water-protein, and appropriate surface charges on the molecules. We selected six representative plasma proteins. As mentioned in Sec.~\ref{sec::protein}, in our CG model each amino acid of a protein is represented by a single bead located at the $\alpha$-carbon position. The native structures are obtained from the Protein Data Bank,
and in Table~\ref{tab::PDB} we list the proteins under study, their PDB IDa from which
the CG model were built and the abbreviations that will be used in the rest of the text.
Table \ref{tab::PDB} also summarizes their relative content in blood and their molar mass.
Although these six proteins represent the most common components of the blood plasma, recent studies\cite{1-36,1-72,1-72a,1-72b} demonstrated that the protein corona can include hundreds of different plasma proteins. Figure~\ref{fig::proteins} shows the CG representations of the six proteins chosen for our study. Note that the range of sizes of the proteins is very broad, from a big molecule as Fib (about $10\times 45$ nm) to a relatively small molecule A1A (about $8 \times 4$ nm). After the CG model were built from the PDB files, the obtained structures were shifted so the COM of the molecules was in the origin of the frame of reference and this structure was defined as the $(\phi=0^{\circ}$,$\theta=0^{\circ})$ orientation. With this definition the
first residue in the sequence of each protein will have the following $(\phi,\theta)$ angles: ($21.4^{\circ}$,$85.2^{\circ}$)
for HSA, ($101.0^{\circ}$,$126.7^{\circ}$) for A1A,
($193.9^{\circ}$,$48.9^{\circ}$) for A2M, ($132.1^{\circ}$,$46.4^{\circ}$) for Fib, ($279.6^{\circ}$,$140.2^{\circ}$)
for Tra and ($6.3^{\circ}$,$110.8^{\circ}$) for IgG.

All simulation were performed using ESPResSo package~\cite{Espresso} and
the cutoff for the interaction potential in Eq.~(\ref{eq::LJRS2}) was set to $r_{\mathrm{cut}}=6$~nm.
For all calculations the simulation box was taken big enough to fit the NP and the protein.
The units of the simulations are: lengths ($\mathcal{L}$) in nm,
energy ($\mathcal{E}$) in $k_B T\approx 4.15 \times10^{-21}~\mathrm{J}$ taking a temperature
of $T=300~\mathrm{K}$, unless specified otherwise. For the mass unit ($\mathcal{M}$) we selected the average
mass of the 20 aminoacids (ca. $110~\mathrm{Da}$) hence in our
simulations all aminoacids have a mass of 1.
The values of $\epsilon_i$ and $\sigma_i$ can be found in Table~\ref{tab::parameters} and
as mentioned in Sec.~\ref{sec::potentials} we will only consider hydrophobic NPs with
$\epsilon_{s}=1$ and $\sigma_s=0.35$~nm.

NPs with negative surface charges with charge density of $-0.05$~C/m$^2$ as well as neutral NPs were considered.
As explained in Sec.~\ref{sec::potentials}, the charged surfaces are modelled by
individual point charges. The surface density of these charged beads ($\sigma_c=N_e/4\pi R^2$)
was set to $\pi^{-1}$~nm$^{-2}$ for all the simulations, which gives \textit{e. g.} a $N_e=100$ for a
NP of $R=5$~nm.
Then, we assumed that each bead carries a charge of $-0.98e$, where $e$ is the elementary charge.
As we are considering physiological conditions, we use $\lambda_B=0.73$~nm and $\lambda_D=1$~nm.
Residue charges at these conditions are $+e$ for LYS and ARG, $-e$ for ASP and GLU, and
$+0.5e$ for HIS. The rest of the residues are neutral.

The only free parameter of the model is $\rho\epsilon_{es}$ in Eq.~(\ref{eq::LJRS}),
and the parametrization was done by systematically changing its
value to match experimental data of adsorption of Lysozyme on
hydrophobic surfaces (octyl- or butyl-sepharose) reported by Chen \textit{et al.}~\cite{CHL2003}.
The native structure for our CG model of Lysozyme was obtained from the PDB ID: 2LYZ.
With $\rho\epsilon_{es}=1.972k_B T/$nm$^3$ we obtain a value of $-7.9 k_B T$ for the
adsorption energy (the same as the experimental reported value).

To validate the parametrization, the adsorption energy of Myoglobin (PDB ID: 1MBN used for the CG model)
was calculated using the same value of $\rho\epsilon_{es}$ obtained from the parametrization.
In this way, a value of $-5.9k_B T$ was found for the adsorption energy of Myoglobin.
This value is slightly lower that the experimental value of $-7.6k_B T$ also reported by Chen \textit{et. al.}~\cite{CHL2003} but
reproduces the trend that Myoglobin adsorbs slightly weaker than Lysozyme to a hydrophobic surface.

\section{Results} \label{sec::results}

\subsection{Mean adsorption energies} \label{sec::Ead}

\begin{figure*}[]
\centering
\includegraphics[width=\hsize]{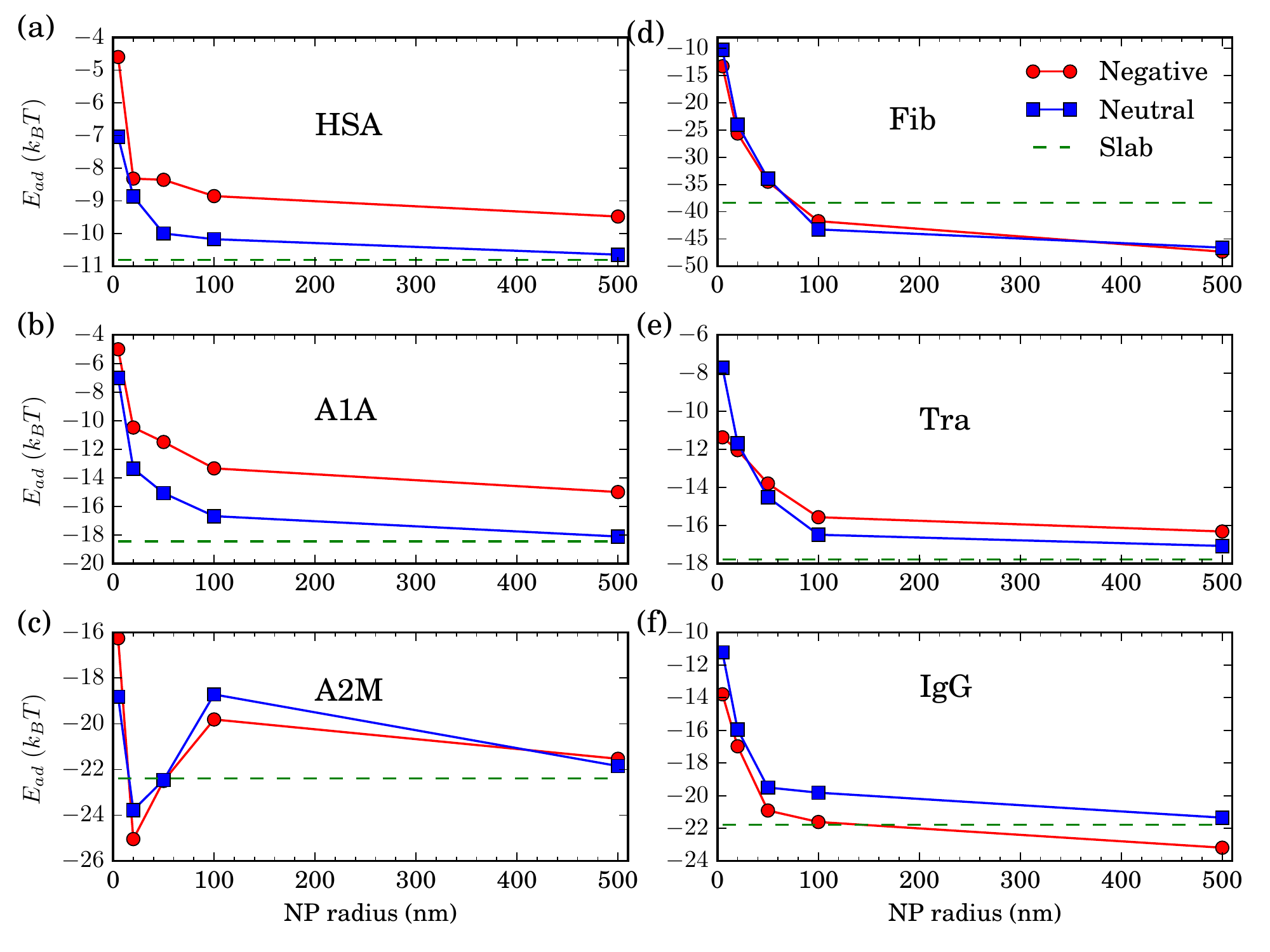}
\caption{
Adsorption energies as a function of the NP radius for the six protein studied and two surface
charge densities: Negative = $-0.05$C/m$^2$ and Neutral = no charge. (A) HSA, (B) A1A, (C) A2M, (D) Fib, (E) Tra and (F) IgG.
The dashed lines show the adsorption energy for the case of a neutral flat surface.}
\label{fig::Eads}
\end{figure*}

Mean adsorption energies for the six proteins calculated using Eq.~(\ref{eq::Esad}) as a function of NP radius are shown in Fig.~\ref{fig::Eads}.
Firstly, the results show that in all cases with the exception of A2M the $E_{ad}$ decreases as the radius of the NP increases.
This trend is due to a combination of two factors: (i) increasing $R$ increases the magnitude of the van der Waals attraction
as shown in Fig.~\ref{fig::potentials} and (ii) increasing the radius of the NP also increases the surface exposed to the proteins.
These effects are more pronounced for NPs of $R<100$ nm, and after this radius the adsorption energies tend to
decrease at a smaller rate. For A2M the curve is non-monotonic and has a minimum of the adsorption energy
for $R=20$~nm. This protein is rather big and has a complex structure, which makes the effects mentioned above
combine in a non-trivial way. We see that for all molecules (except Fib) the total $E_{ad}$ for the neutral
NPs tends to the neutral slab value at $R=500$~nm showing as expected that for large NPs the size has only a small effect
on the van der Waals interactions. For Fib, more points would be needed for $R>500$ nm to observe how the $E_{ad}$ converges
to slab value but our results suggest as in the case of A2M that there is a minimum in the Fib adsorption energy.

Secondly, the effect of the electrostatics is smaller in magnitude than the van der Waals contribution
for the surface charge studied here. This can be confirmed by noticing that the difference between the $E_{ad}$
of the negatively charged and neutral NPs are significantly smaller than $E_{ad}$ itself. In fact,
the electrostatic interactions modify the adsorption energy by less than $3k_B T$ per protein.

Thirdly, Tra and IgG attract stronger to the negatively charged surfaces, HSA and A1A to neutral surfaces while
Fib and A2M do not show a clear pattern. HSA and A1A are sightly negatively charged (both with a total charge
of $-6e$) so overall electrostatic attraction dominates over electrostatic repulsion. For Tra, the
total charge is $+15e$ but excluding the contribution from HIS the total charge is $-4e$. As the HIS residues
contribute half a charge, effectively the positive charge of Tra is less localized and the molecule behaves as a
slightly negative object. To confirm this observation, calculations of the adsorption energy for the
same surfaces but with the HIS with no charge and with charge $+1e$ were done. As expected,
Tra with uncharged HIS residues attach stronger to the a neutral surface, while Tra with
positively charged HIS residues adsorbs stronger onto the negatively charged surface.
In the case of IgG, the total charge of the protein is $+28e$ and stays positive even without the contribution
of the HIS residues (they contribute $+12e$), so the overall electrostatic attraction
is greater for the negatively charged surfaces. Fib is slightly positive ($+3.5e$) so it is expected that it attaches stronger
to the negative surface. This is the case for all radii apart from $R=100$~nm.
Despite that for this radius the Fib molecule attaches stronger to the neutral NP, the difference in the $E_{ad}$
for both surfaces (neutral and charged) is only of about 3.5\%, which again shows that van der Waals contributions dominate
over electrostatic interactions.

For A2M which has a total charge of $-5.5e$, there is no clear indication of the charge effect on $E_{ad}$.
As this molecule is big, the relative contribution of the charge to the adsorption energy is expected to be small compared to the
van der Waals contribution. Our results agree with this prediction as the differences in the $E_{ad}$ are always small
(less than 5\%) with the exception of $R=5$ nm. For this radius, the NP is so small compared to the protein that the contact area includes only few aminoacids.
In this case, the electrostatic interactions have larger relative effect than for the larger sizes.

Lists of the proteins for each NP size, sorted
by the adsorption energy, are reported in Table~\ref{tab::rankneg}.
In all cases, the big proteins (Fib, A2M and IgG) are within the three those most strongly attached to the 
NP independently of the radius, while
the small proteins (A1A, Tra and HSA) are in the group of the three molecules with weaker
adsorption. Also, HSA is the weakest attached protein
and Fib has always the most negative adsorption energy for NPs of $R>5$~nm.

\begin{table}
\centering
\begin{tabular}{ c | c c c c c c }
\hline
\hline
\multicolumn{1}{ c }{ Radius [nm] } & \multicolumn{6}{ c }{ Ranking }\\
\hline
5 & A2M & IgG & Fib & Tra & A1A & HSA \\
20 & Fib & A2M & IgG & Tra & A1A & HSA \\
50 & Fib & A2M & IgG & Tra & A1A & HSA \\
100 & Fib & IgG & A2M & Tra & A1A & HSA \\
500 & Fib & IgG & A2M & Tra & A1A & HSA \\
\hline
\hline
\end{tabular}
\caption{Ranking of the adsorption energies for the negatively charged surface.
For each NP radius, the proteins are sorted from left (stronger adsorption) to right
(weaker adsorption) by their value of $E_{ad}$.}
\label{tab::rankneg}
\end{table}

\subsection{Adsorption energy maps and preferred orientations} \label{sec::orient}

\begin{figure}[]
\centering
\includegraphics[width= \hsize]{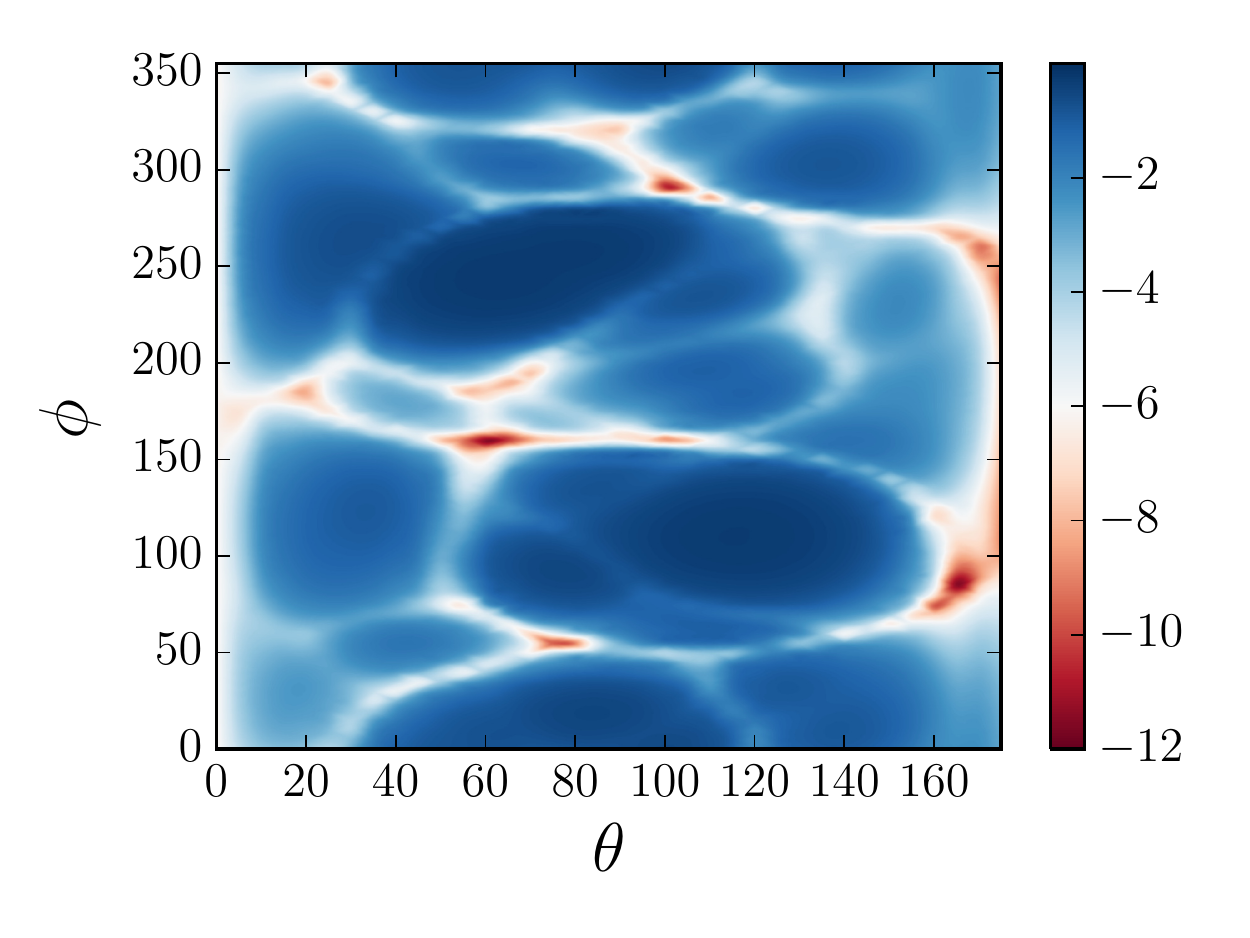}
\caption{Adsorption energy map for HSA interacting with a neutral NP of $R=20$~nm. Energy are is in $k_B T$.}
\label{fig::SMHSAexample}
\end{figure}

The systematic sampling applied for the calculation of the adsorption energies
can also be used to identify the most favorable orientations for the adsorption.
As an example, Fig.~\ref{fig::SMHSAexample} shows a color map of the
adsorption energy as a function of the angles $\theta$ and $\phi$ for HSA adsorbing on a neutral
20~nm-NP. The energy landscape is complex in structure showing several
connected local minima with differences that are less that $1 k_B T$. This observation
suggests that orientational changes in the NP-protein complex after adsorption are likely to occur.
Additionally, we should note that the map has big areas with adsorption energies of $-6 k_B T$ or lower,
which in practice means that more than one orientation gives a relatively strong adsorption, so
that the proteins will bind to NPs at room temperature in various orientations.
For the rest of the proteins and for the conditions studied in this work, similar features are observed
in the energy maps.

\begin{figure*}[]
\centering
\includegraphics[width= \hsize]{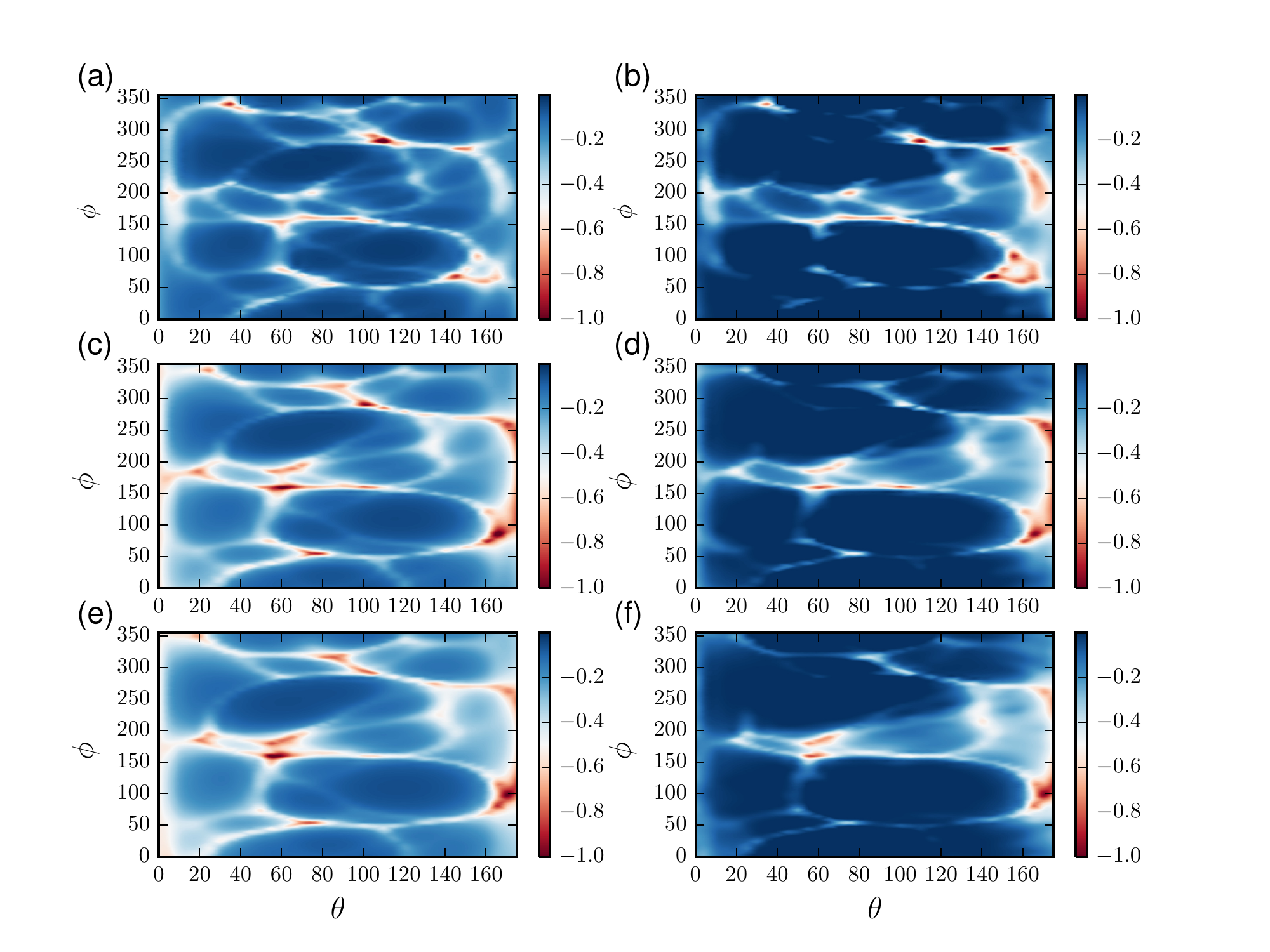}
\caption{Normalized adsorption energy maps for HSA adsorbing on: (a) Neutral NP of $R=5$~nm. (b) Negatively charged NP of $R=5$~nm.
(c) Neutral NP of $R=20$~nm. (d) Negatively charged NP of $R=20$~nm.
(e) Neutral NP of $R=500$~nm. (f) Negatively charged NP of $R=500$~nm.}
\label{fig::SMHSA}
\end{figure*}

To study the effect of the radius of the NP and the charge on the adsorption maps, in Fig.~\ref{fig::SMHSA}
we show the normalized adsorption energy maps for HSA. The adsorption energies have been normalized by rescaling the
energies such that 0 denotes the the maximum adsorption energy while -1 denotes the minimum.
This transformation has been performed to better illustrate the similarities or differences between the different cases.
In Fig.~\ref{fig::SMHSA}, each panel is for a radius of 5, 20 or 500 nm, respectively, and for a neutral or a negative charged surface.
A first comparison of the structure of the maps in the different panels reveals that neither the radius nor the charge
density have a major impact. A closer inspection of the maps for
the same charge but different radii (compare Figs.~\ref{fig::SMHSA}a, \ref{fig::SMHSA}c and \ref{fig::SMHSA}e or Fig.~\ref{fig::SMHSA}b,
\ref{fig::SMHSA}d and \ref{fig::SMHSA}f)
shows that there are only small differences in the structure of the maps. The largest changes
are seen for $R=5$~nm compared to $R=20$ or 500 nm.
On the other hand, the effect of the charge on the structure of the maps
is even smaller (compare Fig.~\ref{fig::SMHSA}a with \ref{fig::SMHSA}b or Fig.~\ref{fig::SMHSA}c with \ref{fig::SMHSA}d
or Figs.~\ref{fig::SMHSA}e with \ref{fig::SMHSA}f).
We performed similar analysis for the other proteins and found that the energy surfaces for Tra and IgG are again very
stable regardless of the surface charge or the radius of the NP.

\begin{figure*}[]
\centering
\includegraphics[width=\hsize]{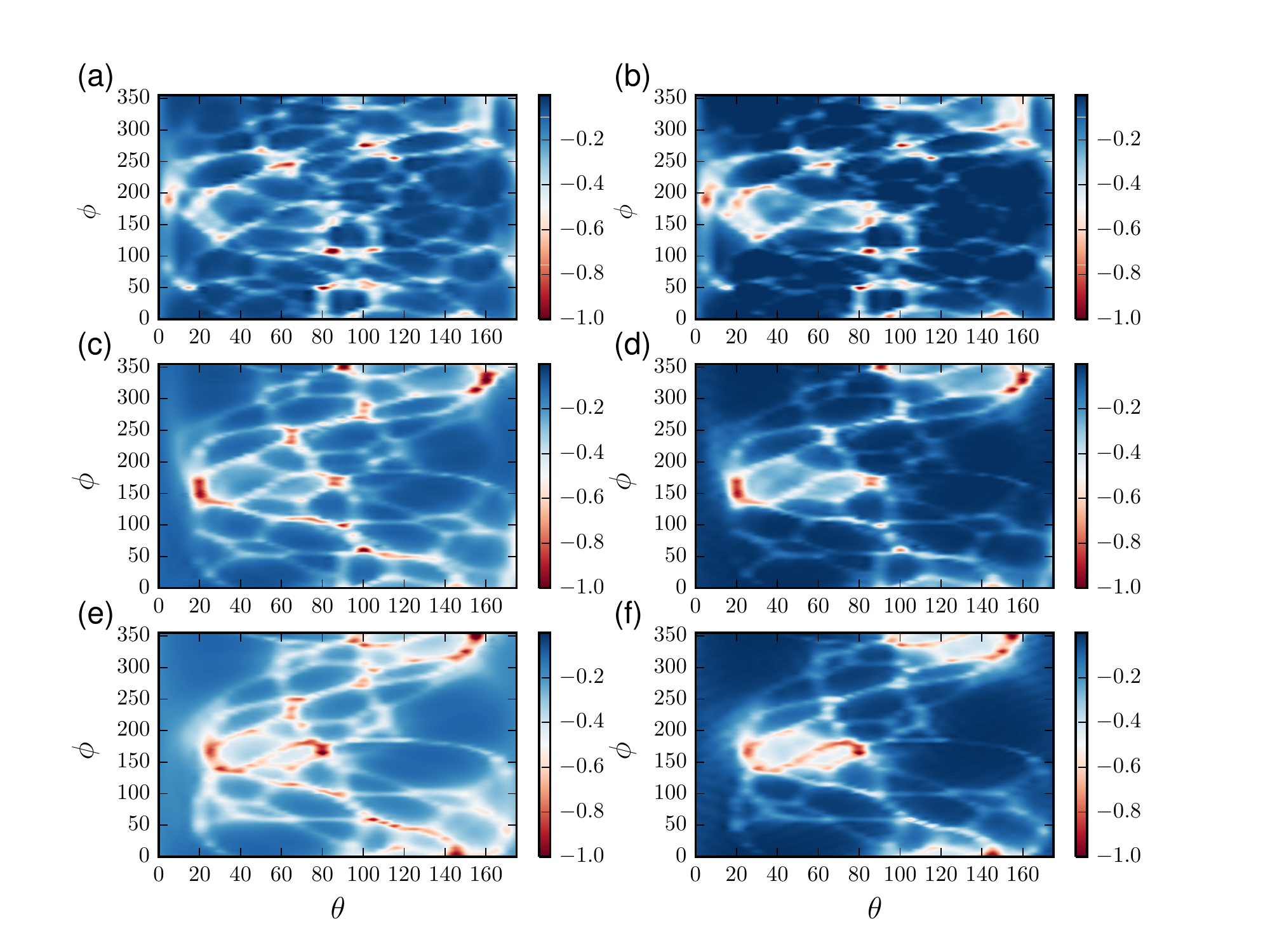}
\caption{Normalized adsorption energy maps for A2M adsorbing on: (a) Neutral NP of $R=5$~nm. (b) Negatively charged NP of $R=5$~nm.
(c) Neutral NP of $R=20$~nm. (d) Negatively charged NP of $R=20$~nm.
(e) Neutral NP of $R=500$~nm. (f) Negatively charged NP of $R=500$~nm.}
\label{fig::SMA2M}
\end{figure*}

A different situation is observed for A2M. Fig~\ref{fig::SMA2M} shows the adsorption energy  map of A2M for
three radii and two charges. The energy maps for $R=5$ nm show clear differences
with the maps for $R=20$~nm and 500~nm. For the smallest radius ($R=5$ nm) the maps contain few local minima compared to the
other two radii. These are also more localized than the local minima observed for $R=20$ and 500 nm.
As in the case of the total adsorption energies, these results can be explained by the size and shape of the big
A2M molecule. For small NPs, the aminoacids that come in contact with the surface in each relative orientation are determined
by specific patch of the protein and as the NPs increases in size these patches become larger allowing
the NP to interact with a larger part of the molecule. The results show that for A2M, a NP of $R=5$~nm is small enough
for the adsorption energy to be affected by the local structure of the molecule but for a $R=20$~nm this effect is already lost.
With respect to the charge, we observe no big difference in the energy maps. As mentioned before, the charge
has a small effect on the total adsorption energy so it is expected that it would not dramatically change the energy maps.

\begin{figure*}[]
\centering
\includegraphics[width= \hsize]{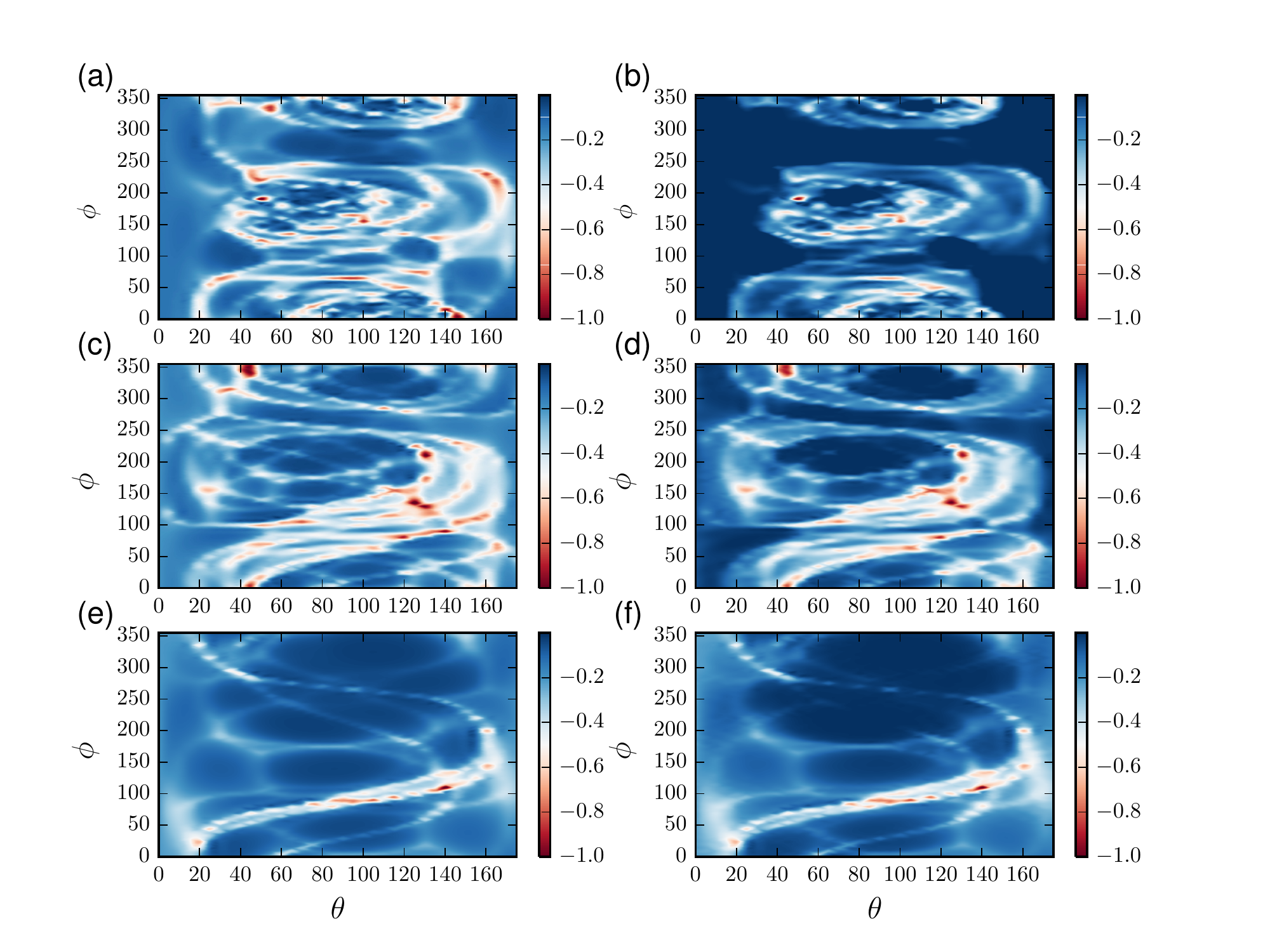}
\caption{Normalized adsorption energy maps for Fib adsorbing on: (a) Neutral NP of $R=5$~nm. (b) Negatively charged NP of $R=5$~nm.
(c) Neutral NP of $R=20$~nm. (d) Negatively charged NP of $R=20$~nm.
(e) Neutral NP of $R=500$~nm. (f) Negatively charged NP of $R=500$~nm.}
\label{fig::SMFib}
\end{figure*}

\begin{figure*}[]
\centering
\includegraphics[width=\hsize]{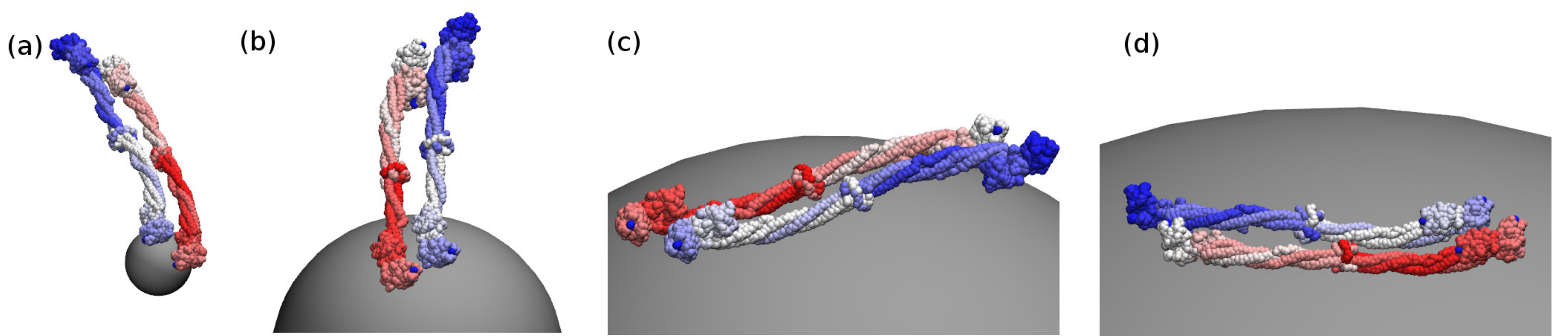}
\caption{Most favorable orientation for the adsorption of Fib on a neutral NP of radius (a) 5, (b) 20, (c) 50 and (d) 100 nm.
The Fib molecule is color coded the same way in all cases.}
\label{fig::Fibscene}
\end{figure*}

For Fib molecule, we observe a different dependence of the energy maps on the NP size.
Fig.~\ref{fig::SMFib} shows the energy landscapes obtained for NPs of radii 5, 20 and 500~nm for the neutral
and charged surfaces. In this case, difference can be appreciated from comparing the maps for the three radii
(compare Fig.~\ref{fig::SMFib} \ref{fig::SMFib}a, \ref{fig::SMFib}c and \ref{fig::SMFib}e or
Fig.~\ref{fig::SMFib}b, \ref{fig::SMFib}d and \ref{fig::SMFib}f). Fib is not only a big molecule
but it is also long, so the size effect explained above for A1M is enhanced. The energies for Fib calculated
for the other NP radius used in this work indicate that for a $R>50$~nm the maps do not
contain noticeable differences (results not shown). To better illustrate this result, in Fig.~\ref{fig::Fibscene}
we show the most favorable orientations for Fib on a neutral surface for four different NP radii ($R=5$, 20, 50 and 100 nm).
For the two smallest NPs (Fig.~\ref{fig::Fibscene}a and \ref{fig::Fibscene}b), Fib has its adsorption energy minimum in a configuration where
the NP is attached to a tip of the molecule. The main difference between the $R=5$~nm and 20~nm NP-protein complexes
is that the second one is interacting with a larger portion of the tip. Meanwhile, as the NP increases in size,
Fib tends to adsorb in a sidewise orientation (Fig.~\ref{fig::Fibscene}c and \ref{fig::Fibscene}d).
These means that the most preferred orientation is the one that corresponds to the longest axis
of the Fib molecule stretched along the surface, which maximizes the number of the aminoacids in direct contact with the NP.
As for the other proteins, the surface charge of the NP does not induce noticeable differences
(compare Fig.~\ref{fig::SMFib}a with \ref{fig::SMFib}b or Fig.~\ref{fig::SMFib}c
with \ref{fig::SMFib}d or Fig.~\ref{fig::SMFib}e with \ref{fig::SMFib}f).
For Tra molecule, we find similar behavior for the dependence of the energy landscape on the NP size to
that observed for Fib, with the only difference that the landscapes do not change anymore for $R>20$~nm.
The effect is again due to the size of the molecule, which is not as big as Fib, although elongated and thus
is distinct from the more spherical A1M or IgG.

\subsection{Effect of temperature on the adsorption energy} \label{sec::Teffect}

The results reported until now were obtained at $T=300$~K, which is a temperature commonly used in
\textit{in vitro} experiments. Now we also considere the adsorption of HSA at a temperature of $T=310$~K, 
which is more relevant for \emph{in vivo} conditions. For the calculations
at $T=300$~K the value of the free parameter of the model ($\rho\epsilon_{es}=1.972k_B T/$nm$^3$) was
obtained from matching the $E_{ad}$ calculated using Eq.~(\ref{eq::Esad}) to the value of $-7.9 k_B T$
reported by Chen {\it et al.}~\cite{CHL2003} as the adsorption energy of Lysozyme on hydrophobic surfaces at
$T=300$ K (for more details on the parametrization, see Sec.~\ref{sec::Para}). In the same experimental
work, a value of $-8.2 k_B T$ for adsorption energy of Lysozyme on a hydrophobic surfaces at
$T=310$ K is given which we used to scale the free parameter of the model. In this way, for the simulations
at $T=310$ K a value of $\rho\epsilon_{es}=2.07k_B T/$~nm$^3$ was used. Additionally, the parameters for the
electrostatic interactions were also changed to account for a temperature of 310 K:
$\lambda_B=0.72$~nm and $\lambda_D=0.96$~nm.

\begin{figure}[]
\centering
\includegraphics[width=\hsize]{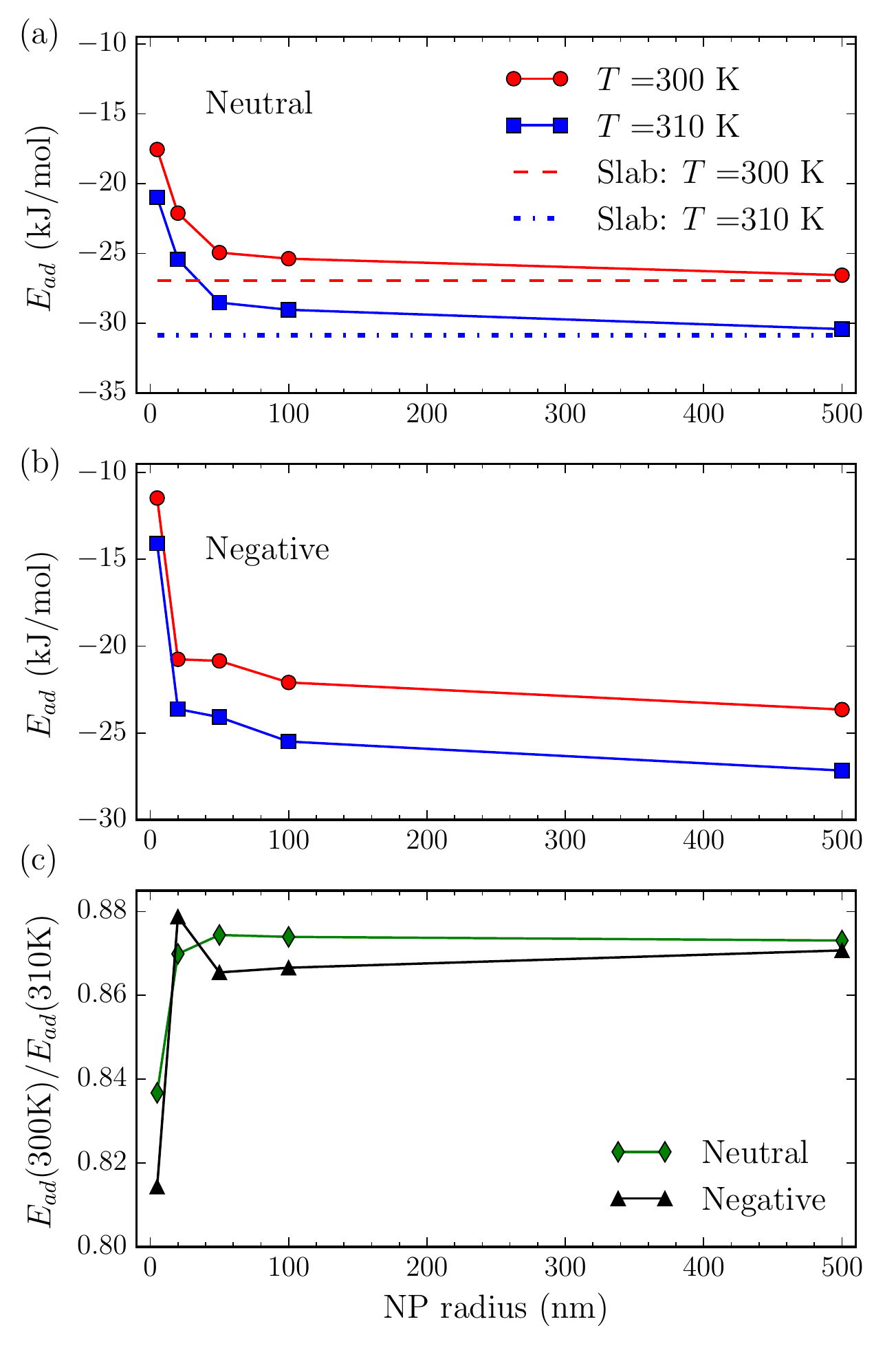}
\caption{Adsorption energies for HSA as a function of the NP radius for two different temperatures.
(a) Adsorption energies for neutral NPs at $T=300$~K (circles) and $T=310$~K (squares).
(b) Adsorption energies for negatively charged NPs at $T=300$~K (circles) and $T=310$~K (squares).
The dashed lines are the values of $E_{ad}$ for the slab cases at $T=300$~K (red) and $T=300$~K (blue).
(c) The ratio $\gamma=E_{ad}(T=300 \mathrm{K})/E_{ad}(T=310 \mathrm{K})$ for neutral (diamonds)
and negatively charged (triangles) NPs.}
\label{fig::EadsT}
\end{figure}

Figs.~\ref{fig::EadsT}a and Figs.~\ref{fig::EadsT}b show the adsorption energy for HSA as a function of $R$
at the two studied temperatures for the neutral and negatively charged surfaces, respectively.
As expected, the HSA molecule attaches stronger at $T=310$ K to both surfaces (neutral and negative).
More importantly, the temperature does not change substantially the shapes of the curves, which
suggests that the effect of the temperature increase  simply shifts the adsorption energy toward the more negative values.
As mentioned in the previous sections, the van der Waals interactions dominate over electrostatics and
as they depend linearly on the parameter $\rho\epsilon_{es}$ the overall effect merely reflects this trend. 
To get a better insight on how this energy change depends on the radius of the NP,
in Fig.~\ref{fig::EadsT}c the ratio between the adsorption energies at the two temperatures is shown.
For the smaller NP ($R=5$~nm) the ratio is lower than for the other cases, indicating
that for this radius the effect of the temperature on the adsorption energy is more pronounced.
For large NPs, the ratio is greater and tends to a value of about 0.88, which is less
than the ratio between the energy scaling parameters $\rho\epsilon_{es}$'s for $T=300$~K over 
$T=310$~K of 0.92.
Thus, we see that the effect of the temperature on the adsorption energies is not only in the uniform decrease of the energy, which
is not very surprising as $E_{ad}$ for each orientation defined through an sophisticated integral (see Eq.~(\ref{eq::Esad1})), so modulating
the value of $\rho\epsilon_{es}$ will affect the overall $E_{ad}$ in a non-linear way.
As for $T=300$ K, the charge of the surface at 310 K has a small effect on the adsorption
energies. Finally, we compared the adsorption energy landscapes for the two temperatures (results not shown) and found that they
do not differ from each other.

\section{Discussion} \label{sec::discussion}

We can now validate the CG model by comparing the predictions with known simulation and
experimental data. Lacerda~\textit{et. al}~\cite{LPJ2010} measured the binding association constant
for HSA, Fib and a set of $\gamma$-globulins on citrate-coated gold
NPs (which can be considered as negative moderately hydrophobic NPs). They find
that for NPs with $R=50$~nm (this was the biggest in their study),
the $\gamma$-globulins are the proteins that adsorb most strongly, followed
by Fib and finishing with HSA. Our rankings reported in Table~\ref{tab::rankneg}
are slightly different as Fib is predicted to attach stronger that IgG.
For NPs of $15<R<30$~nm, the experiments show that HSA is still the weakest
adsorbing molecule, but Fib shows equal or largest
binding association constant than the $\gamma$-globulins, which agrees with our results.
For $R=5$~nm, the experiments show again that HSA has the smallest
binding association constant while Fib and the $\gamma$-globulins exhibit the same
affinity. This again fits well with our finding as Fib and IgG have a very similar
$E_{ad}$ for $R=5$~nm ($-13.8 k_B T$ for IgG and $-13.3 k_B T$ for Fib).
This discrepancy for the bigger radius are mainly due to that in are calculations we are considering only one
$\gamma$-globulin while the experimental data was collected for a set of proteins of similar
structure. It is also interesting to compare our results with the simulations
reported by Vilaseca \textit{et al.},\cite{VDF2013} who studied competitive adsorption of proteins
on surfaces. Using CG MD simulations, they found that for a flat surface at long times the most abundant
protein adsorbed were Fib, then IgG and at last HSA, which also agrees with the ranking
based on adsorption energies. Generally, this phenomenon of adsorbed protein replacement is known in 
literature as the Vroman effect.\cite{Vroman1962,DeLuc1995,Ortega1995,Holmberg2009,Oberle2015}
In general, we find bigger proteins adsorb stronger on the surfaces, even for small NPs.

It is important to remark that the adsorption energies and rankings calculated in this work
can help to predict the long-time composition of the NP-protein corona
but at short times other factors such as the protein sizes, their concentrations and mobilities
have to be considered. As calculations for any protein with known crystal structure can be done easily in 
our CG model, its results can be used as an input for studying competitive adsorption
of proteins and the corona formation kinetics, such as the models in Refs.\cite{VDF2013,Oberle2015}

\begin{figure}[]
\centering
\includegraphics[width=\hsize]{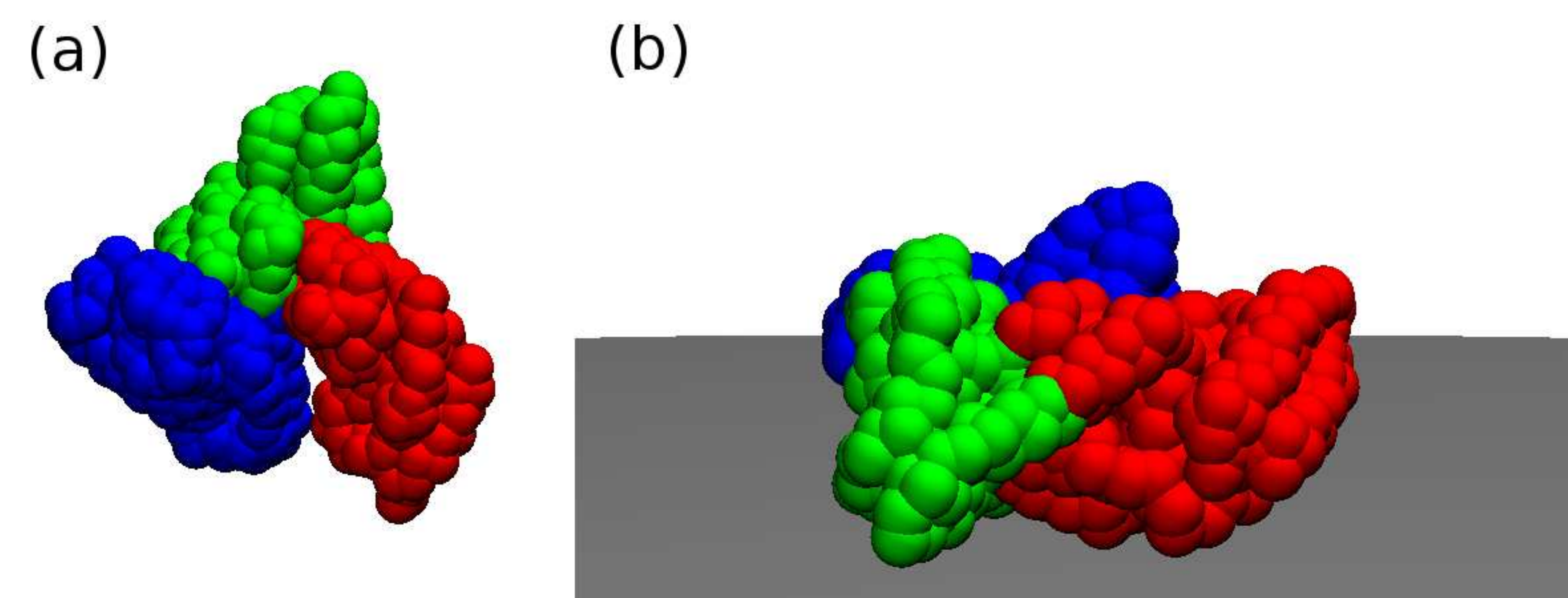}
\caption{(a) Domains of HSA molecule.
(b) Most favourable orientation of HSA on a flat hydrophobic surface.}
\label{fig::HSAconfi}
\end{figure}

In Fig. \ref{fig::HSAconfi} we show a color-coded representation of HSA divided in three domains and the most favorable orientation for adsorption on a $R = 500$~nm NP (for HSA at this radius the NP the adsorption preference is equivalent to that on a flat surface). The HSA molecule adsorbs in a side-on configuration, in which all three domains are interacting with the surface. Khan \emph{et al.},\cite{KGN2013} have reported recently a similar result for the adsorption of HSA on a hydrophobic surface but using docking simulations based on full atomistic model. We conclude that our CG scheme indeed preserves enough information to capture the essential adsorption mechanisms. Therefore, our model can be used for a quick search for the preferred configurations to accelerate docking experiments or to study competitive protein adsorption in plasma or other protein solutions.\cite{Oberle2015}

The main limitation of our model is that as we consider the proteins to be rigid bodies, conformational changes are not allowed and in some cases this can be an important factor for the adsorption process.\cite{LPJ2010,Schmid2014} This assumption can be relaxed by \textit{e.g.} using elastic network derived from the principal component analysis of the molecule dynamics\cite{Schmid2014} or a G\={o}--Type model (see~\cite{Noid2013} for a review on CG models of proteins). Another limitation is the inability to address properly the interaction of hydrated residues/surfaces. This deficiency can be corrected by introduction of more accurate potentials of mean force that take into account the water and surface structuring, which can be derived from all-atom MD simulations.\cite{Brandt2014,Brandt2015} We are currently working on combining the PMFs from atomistic MD simulations with our model of proteins.

\section{Conclusions}\label{sec::conclusions}

In this work, we presented a CG model for calculation of the adsorption energies of globular proteins on hydrophobic charged NPs. The proposed method was
parameterized and validated against more detailed simulation and experiments. The model can be applied for evaluation of binding energies for
arbitrary plasma, cytosolic or membrane proteins with known structure, ranking them by binding affinity to the NP and predicting the content
of NP protein corona. We performed a study of adsorption of six common blood plasma proteins onto hydrophobic NPs. We found that the NP surface charge 
has a small effect on the adsorption energies in comparison to van der Waals interactions between the residues and the surface. We also found that
the charge of the NP does not have a major influence on the orientation, in which the proteins are to be mostly likely adsorbed.
On the other hand, we showed that the size of the NP has a pronounced effect on the adsorption energy maps, as the
curvature of the NP determine the sections of the protein that can come in contact with the NP surface. 
Finally, we note that, as the methodology presented here is computationally efficient and gives consistent predictions for the structure of 
NP-protein complexes, it can be used as a part of multiscale modelling methodologies or a supplement to more sophisticated approaches to 
modelling bionano interactions.

\begin{acknowledgments}
The presented research has been funded by EU FP7 collaborative grant, project 310465 (MembraneNanoPart).
\end{acknowledgments}

\bibliographystyle{aipnum4-1}
\bibliography{refs}

\begin{thebibliography}{46}%
\makeatletter
\providecommand \@ifxundefined [1]{%
 \@ifx{#1\undefined}
}%
\providecommand \@ifnum [1]{%
 \ifnum #1\expandafter \@firstoftwo
 \else \expandafter \@secondoftwo
 \fi
}%
\providecommand \@ifx [1]{%
 \ifx #1\expandafter \@firstoftwo
 \else \expandafter \@secondoftwo
 \fi
}%
\providecommand \natexlab [1]{#1}%
\providecommand \enquote  [1]{``#1''}%
\providecommand \bibnamefont  [1]{#1}%
\providecommand \bibfnamefont [1]{#1}%
\providecommand \citenamefont [1]{#1}%
\providecommand \href@noop [0]{\@secondoftwo}%
\providecommand \href [0]{\begingroup \@sanitize@url \@href}%
\providecommand \@href[1]{\@@startlink{#1}\@@href}%
\providecommand \@@href[1]{\endgroup#1\@@endlink}%
\providecommand \@sanitize@url [0]{\catcode `\\12\catcode `\$12\catcode
  `\&12\catcode `\#12\catcode `\^12\catcode `\_12\catcode `\%12\relax}%
\providecommand \@@startlink[1]{}%
\providecommand \@@endlink[0]{}%
\providecommand \url  [0]{\begingroup\@sanitize@url \@url }%
\providecommand \@url [1]{\endgroup\@href {#1}{\urlprefix }}%
\providecommand \urlprefix  [0]{URL }%
\providecommand \Eprint [0]{\href }%
\providecommand \doibase [0]{http://dx.doi.org/}%
\providecommand \selectlanguage [0]{\@gobble}%
\providecommand \bibinfo  [0]{\@secondoftwo}%
\providecommand \bibfield  [0]{\@secondoftwo}%
\providecommand \translation [1]{[#1]}%
\providecommand \BibitemOpen [0]{}%
\providecommand \bibitemStop [0]{}%
\providecommand \bibitemNoStop [0]{.\EOS\space}%
\providecommand \EOS [0]{\spacefactor3000\relax}%
\providecommand \BibitemShut  [1]{\csname bibitem#1\endcsname}%
\let\auto@bib@innerbib\@empty
\bibitem [{\citenamefont {Monopoli}\ \emph {et~al.}(2012)\citenamefont
  {Monopoli}, \citenamefont {Aberg}, \citenamefont {Salvati},\ and\
  \citenamefont {Dawson}}]{MASD2012}%
  \BibitemOpen
  \bibfield  {author} {\bibinfo {author} {\bibfnamefont {M.}~\bibnamefont
  {Monopoli}}, \bibinfo {author} {\bibfnamefont {C.}~\bibnamefont {Aberg}},
  \bibinfo {author} {\bibfnamefont {A.}~\bibnamefont {Salvati}}, \ and\
  \bibinfo {author} {\bibfnamefont {K.~A.}\ \bibnamefont {Dawson}},\ }\href
  {\doibase 10.1038/NNANO.2012.207} {\bibfield  {journal} {\bibinfo  {journal}
  {Nat. Nanotechnol.}\ }\textbf {\bibinfo {volume} {7}},\ \bibinfo {pages}
  {779} (\bibinfo {year} {2012})}\BibitemShut {NoStop}%
\bibitem [{\citenamefont {Lynch}, \citenamefont {Dawson},\ and\ \citenamefont
  {Linse}(2006)}]{1-72}%
  \BibitemOpen
  \bibfield  {author} {\bibinfo {author} {\bibfnamefont {I.}~\bibnamefont
  {Lynch}}, \bibinfo {author} {\bibfnamefont {K.~A.}\ \bibnamefont {Dawson}}, \
  and\ \bibinfo {author} {\bibfnamefont {S.}~\bibnamefont {Linse}},\ }\href
  {\doibase 10.1126/stke.3272006pe14} {\bibfield  {journal} {\bibinfo
  {journal} {Sci. Signal.}\ }\textbf {\bibinfo {volume} {2006}},\ \bibinfo
  {pages} {pe14} (\bibinfo {year} {2006})}\BibitemShut {NoStop}%
\bibitem [{\citenamefont {Cedervall}\ \emph {et~al.}(2007)\citenamefont
  {Cedervall}, \citenamefont {Lynch}, \citenamefont {Lindman}, \citenamefont
  {Berggard}, \citenamefont {Thulin}, \citenamefont {Nilsson}, \citenamefont
  {Dawson},\ and\ \citenamefont {Linse}}]{1-72a}%
  \BibitemOpen
  \bibfield  {author} {\bibinfo {author} {\bibfnamefont {T.}~\bibnamefont
  {Cedervall}}, \bibinfo {author} {\bibfnamefont {I.}~\bibnamefont {Lynch}},
  \bibinfo {author} {\bibfnamefont {S.}~\bibnamefont {Lindman}}, \bibinfo
  {author} {\bibfnamefont {T.}~\bibnamefont {Berggard}}, \bibinfo {author}
  {\bibfnamefont {E.}~\bibnamefont {Thulin}}, \bibinfo {author} {\bibfnamefont
  {H.}~\bibnamefont {Nilsson}}, \bibinfo {author} {\bibfnamefont {K.~A.}\
  \bibnamefont {Dawson}}, \ and\ \bibinfo {author} {\bibfnamefont
  {S.}~\bibnamefont {Linse}},\ }\href@noop {} {\bibfield  {journal} {\bibinfo
  {journal} {Proc. Natl. Acad. Sci. U.S.A.}\ }\textbf {\bibinfo {volume}
  {104}},\ \bibinfo {pages} {2050} (\bibinfo {year} {2007})}\BibitemShut
  {NoStop}%
\bibitem [{\citenamefont {Lindman}\ \emph {et~al.}(2007)\citenamefont
  {Lindman}, \citenamefont {Lynch}, \citenamefont {Thulin}, \citenamefont
  {Nilsson}, \citenamefont {Dawson},\ and\ \citenamefont {Linse}}]{1-72b}%
  \BibitemOpen
  \bibfield  {author} {\bibinfo {author} {\bibfnamefont {S.}~\bibnamefont
  {Lindman}}, \bibinfo {author} {\bibfnamefont {I.}~\bibnamefont {Lynch}},
  \bibinfo {author} {\bibfnamefont {E.}~\bibnamefont {Thulin}}, \bibinfo
  {author} {\bibfnamefont {H.}~\bibnamefont {Nilsson}}, \bibinfo {author}
  {\bibfnamefont {K.~A.}\ \bibnamefont {Dawson}}, \ and\ \bibinfo {author}
  {\bibfnamefont {S.}~\bibnamefont {Linse}},\ }\href@noop {} {\bibfield
  {journal} {\bibinfo  {journal} {Nano Lett.}\ }\textbf {\bibinfo {volume}
  {7}},\ \bibinfo {pages} {914} (\bibinfo {year} {2007})}\BibitemShut {NoStop}%
\bibitem [{\citenamefont {Allen}\ \emph {et~al.}(2006)\citenamefont {Allen},
  \citenamefont {Tosetto}, \citenamefont {Miller}, \citenamefont {O'Connor},
  \citenamefont {Penney}, \citenamefont {Lynch}, \citenamefont {Keenan},
  \citenamefont {Pennington}, \citenamefont {Dawson},\ and\ \citenamefont
  {Gallagher}}]{1-72c}%
  \BibitemOpen
  \bibfield  {author} {\bibinfo {author} {\bibfnamefont {L.}~\bibnamefont
  {Allen}}, \bibinfo {author} {\bibfnamefont {M.}~\bibnamefont {Tosetto}},
  \bibinfo {author} {\bibfnamefont {I.}~\bibnamefont {Miller}}, \bibinfo
  {author} {\bibfnamefont {D.}~\bibnamefont {O'Connor}}, \bibinfo {author}
  {\bibfnamefont {S.}~\bibnamefont {Penney}}, \bibinfo {author} {\bibfnamefont
  {I.}~\bibnamefont {Lynch}}, \bibinfo {author} {\bibfnamefont
  {A.}~\bibnamefont {Keenan}}, \bibinfo {author} {\bibfnamefont
  {S.}~\bibnamefont {Pennington}}, \bibinfo {author} {\bibfnamefont
  {K.}~\bibnamefont {Dawson}}, \ and\ \bibinfo {author} {\bibfnamefont
  {W.}~\bibnamefont {Gallagher}},\ }\href@noop {} {\bibfield  {journal}
  {\bibinfo  {journal} {Biomaterials}\ }\textbf {\bibinfo {volume} {27}},\
  \bibinfo {pages} {3096} (\bibinfo {year} {2006})}\BibitemShut {NoStop}%
\bibitem [{\citenamefont {Kamath}\ \emph {et~al.}(2015)\citenamefont {Kamath},
  \citenamefont {Fernandez}, \citenamefont {Giralt},\ and\ \citenamefont
  {Rallo}}]{Kamath2015}%
  \BibitemOpen
  \bibfield  {author} {\bibinfo {author} {\bibfnamefont {P.}~\bibnamefont
  {Kamath}}, \bibinfo {author} {\bibfnamefont {A.}~\bibnamefont {Fernandez}},
  \bibinfo {author} {\bibfnamefont {F.}~\bibnamefont {Giralt}}, \ and\ \bibinfo
  {author} {\bibfnamefont {R.}~\bibnamefont {Rallo}},\ }\href@noop {}
  {\bibfield  {journal} {\bibinfo  {journal} {Curr. Top. Med. Chem.}\ }\textbf
  {\bibinfo {volume} {15}},\ \bibinfo {pages} {1930} (\bibinfo {year}
  {2015})}\BibitemShut {NoStop}%
\bibitem [{\citenamefont {Salata}(2004)}]{Salata2004}%
  \BibitemOpen
  \bibfield  {author} {\bibinfo {author} {\bibfnamefont {O.~V.}\ \bibnamefont
  {Salata}},\ }\href {\doibase 10.1186/1477-3155-2-3} {\bibfield  {journal}
  {\bibinfo  {journal} {J. Nanobiotechnol.}\ }\textbf {\bibinfo {volume} {2}},\
  \bibinfo {pages} {3} (\bibinfo {year} {2004})}\BibitemShut {NoStop}%
\bibitem [{\citenamefont {Rahman}\ \emph {et~al.}(2012)\citenamefont {Rahman},
  \citenamefont {Ahmad}, \citenamefont {Kazmi}, \citenamefont {Akhter},
  \citenamefont {Afzal}, \citenamefont {Gupta},\ and\ \citenamefont
  {V.R.}}]{Rahman2012}%
  \BibitemOpen
  \bibfield  {author} {\bibinfo {author} {\bibfnamefont {M.}~\bibnamefont
  {Rahman}}, \bibinfo {author} {\bibfnamefont {M.}~\bibnamefont {Ahmad}},
  \bibinfo {author} {\bibfnamefont {I.}~\bibnamefont {Kazmi}}, \bibinfo
  {author} {\bibfnamefont {S.}~\bibnamefont {Akhter}}, \bibinfo {author}
  {\bibfnamefont {M.}~\bibnamefont {Afzal}}, \bibinfo {author} {\bibfnamefont
  {G.}~\bibnamefont {Gupta}}, \ and\ \bibinfo {author} {\bibfnamefont
  {S.}~\bibnamefont {V.R.}},\ }\href@noop {} {\bibfield  {journal} {\bibinfo
  {journal} {Curr Drug Discov Technol.}\ }\textbf {\bibinfo {volume} {9}},\
  \bibinfo {pages} {319} (\bibinfo {year} {2012})}\BibitemShut {NoStop}%
\bibitem [{\citenamefont {Wang}, \citenamefont {Niu},\ and\ \citenamefont
  {Chen}(2014)}]{Wang2014}%
  \BibitemOpen
  \bibfield  {author} {\bibinfo {author} {\bibfnamefont {Z.}~\bibnamefont
  {Wang}}, \bibinfo {author} {\bibfnamefont {G.}~\bibnamefont {Niu}}, \ and\
  \bibinfo {author} {\bibfnamefont {X.}~\bibnamefont {Chen}},\ }\href {\doibase
  10.1007/s11095-013-1103-7} {\bibfield  {journal} {\bibinfo  {journal} {Pharm
  Res.}\ }\textbf {\bibinfo {volume} {31}},\ \bibinfo {pages} {1358} (\bibinfo
  {year} {2014})}\BibitemShut {NoStop}%
\bibitem [{\citenamefont {Roecker}\ \emph {et~al.}(2009)\citenamefont
  {Roecker}, \citenamefont {Poetzl}, \citenamefont {Zhang}, \citenamefont
  {Parak},\ and\ \citenamefont {Nienhaus}}]{Roeckeretaal2009}%
  \BibitemOpen
  \bibfield  {author} {\bibinfo {author} {\bibfnamefont {C.}~\bibnamefont
  {Roecker}}, \bibinfo {author} {\bibfnamefont {M.}~\bibnamefont {Poetzl}},
  \bibinfo {author} {\bibfnamefont {F.}~\bibnamefont {Zhang}}, \bibinfo
  {author} {\bibfnamefont {W.~J.}\ \bibnamefont {Parak}}, \ and\ \bibinfo
  {author} {\bibfnamefont {G.~U.}\ \bibnamefont {Nienhaus}},\ }\href {\doibase
  10.1038/NNANO.2009.195} {\bibfield  {journal} {\bibinfo  {journal} {Nature
  Nanotech.}\ }\textbf {\bibinfo {volume} {4}},\ \bibinfo {pages} {577}
  (\bibinfo {year} {2009})}\BibitemShut {NoStop}%
\bibitem [{\citenamefont {Kelly}\ \emph {et~al.}(2015)\citenamefont {Kelly},
  \citenamefont {Aberg}, \citenamefont {Polo}, \citenamefont {O'Connell},
  \citenamefont {Cookman}, \citenamefont {Fallon}, \citenamefont {Krpetic},\
  and\ \citenamefont {Dawson}}]{Keatal2015}%
  \BibitemOpen
  \bibfield  {author} {\bibinfo {author} {\bibfnamefont {P.~M.}\ \bibnamefont
  {Kelly}}, \bibinfo {author} {\bibfnamefont {C.}~\bibnamefont {Aberg}},
  \bibinfo {author} {\bibfnamefont {E.}~\bibnamefont {Polo}}, \bibinfo {author}
  {\bibfnamefont {A.}~\bibnamefont {O'Connell}}, \bibinfo {author}
  {\bibfnamefont {J.}~\bibnamefont {Cookman}}, \bibinfo {author} {\bibfnamefont
  {J.}~\bibnamefont {Fallon}}, \bibinfo {author} {\bibfnamefont
  {Z.}~\bibnamefont {Krpetic}}, \ and\ \bibinfo {author} {\bibfnamefont
  {K.~A.}\ \bibnamefont {Dawson}},\ }\href {\doibase 10.1038/NNANO.2015.47}
  {\bibfield  {journal} {\bibinfo  {journal} {Nature Nanotech.}\ }\textbf
  {\bibinfo {volume} {10}},\ \bibinfo {pages} {472} (\bibinfo {year}
  {2015})}\BibitemShut {NoStop}%
\bibitem [{\citenamefont {Silva}\ \emph {et~al.}(2006)\citenamefont {Silva},
  \citenamefont {Gorenstein}, \citenamefont {Li}, \citenamefont {Vissers},\
  and\ \citenamefont {Geromanos}}]{Silva2006}%
  \BibitemOpen
  \bibfield  {author} {\bibinfo {author} {\bibfnamefont {J.~C.}\ \bibnamefont
  {Silva}}, \bibinfo {author} {\bibfnamefont {M.~V.}\ \bibnamefont
  {Gorenstein}}, \bibinfo {author} {\bibfnamefont {G.-Z.}\ \bibnamefont {Li}},
  \bibinfo {author} {\bibfnamefont {J.~P.~C.}\ \bibnamefont {Vissers}}, \ and\
  \bibinfo {author} {\bibfnamefont {S.~J.}\ \bibnamefont {Geromanos}},\ }\href
  {\doibase 10.1074/mcp.M500230-MCP200} {\bibfield  {journal} {\bibinfo
  {journal} {Mol. Cell. Proteomics}\ }\textbf {\bibinfo {volume} {5}},\
  \bibinfo {pages} {144} (\bibinfo {year} {2006})}\BibitemShut {NoStop}%
\bibitem [{\citenamefont {Ritz}\ \emph {et~al.}(2015)\citenamefont {Ritz},
  \citenamefont {Schoettler}, \citenamefont {Kotman}, \citenamefont {Baier},
  \citenamefont {Musyanovych}, \citenamefont {Kuharev}, \citenamefont
  {Landfester}, \citenamefont {Schild}, \citenamefont {Jahn}, \citenamefont
  {Tenzer},\ and\ \citenamefont {Mailaender}}]{Ritz2015}%
  \BibitemOpen
  \bibfield  {author} {\bibinfo {author} {\bibfnamefont {S.}~\bibnamefont
  {Ritz}}, \bibinfo {author} {\bibfnamefont {S.}~\bibnamefont {Schoettler}},
  \bibinfo {author} {\bibfnamefont {N.}~\bibnamefont {Kotman}}, \bibinfo
  {author} {\bibfnamefont {G.}~\bibnamefont {Baier}}, \bibinfo {author}
  {\bibfnamefont {A.}~\bibnamefont {Musyanovych}}, \bibinfo {author}
  {\bibfnamefont {J.}~\bibnamefont {Kuharev}}, \bibinfo {author} {\bibfnamefont
  {K.}~\bibnamefont {Landfester}}, \bibinfo {author} {\bibfnamefont
  {H.}~\bibnamefont {Schild}}, \bibinfo {author} {\bibfnamefont
  {O.}~\bibnamefont {Jahn}}, \bibinfo {author} {\bibfnamefont {S.}~\bibnamefont
  {Tenzer}}, \ and\ \bibinfo {author} {\bibfnamefont {V.}~\bibnamefont
  {Mailaender}},\ }\href {\doibase 10.1021/acs.biomac.5b00108} {\bibfield
  {journal} {\bibinfo  {journal} {Biomacromolecules}\ }\textbf {\bibinfo
  {volume} {16}},\ \bibinfo {pages} {1311} (\bibinfo {year}
  {2015})}\BibitemShut {NoStop}%
\bibitem [{\citenamefont {Winzen}\ \emph {et~al.}(2015)\citenamefont {Winzen},
  \citenamefont {Schoettler}, \citenamefont {Baier}, \citenamefont {Rosenauer},
  \citenamefont {Mailaender}, \citenamefont {Landfester},\ and\ \citenamefont
  {Mohr}}]{Winzen2015}%
  \BibitemOpen
  \bibfield  {author} {\bibinfo {author} {\bibfnamefont {S.}~\bibnamefont
  {Winzen}}, \bibinfo {author} {\bibfnamefont {S.}~\bibnamefont {Schoettler}},
  \bibinfo {author} {\bibfnamefont {G.}~\bibnamefont {Baier}}, \bibinfo
  {author} {\bibfnamefont {C.}~\bibnamefont {Rosenauer}}, \bibinfo {author}
  {\bibfnamefont {V.}~\bibnamefont {Mailaender}}, \bibinfo {author}
  {\bibfnamefont {K.}~\bibnamefont {Landfester}}, \ and\ \bibinfo {author}
  {\bibfnamefont {K.}~\bibnamefont {Mohr}},\ }\href {\doibase
  10.1039/C4NR05982D} {\bibfield  {journal} {\bibinfo  {journal} {Nanoscale}\
  }\textbf {\bibinfo {volume} {7}},\ \bibinfo {pages} {2992} (\bibinfo {year}
  {2015})}\BibitemShut {NoStop}%
\bibitem [{\citenamefont {del Pino}\ \emph {et~al.}(2014)\citenamefont {del
  Pino}, \citenamefont {Pelaz}, \citenamefont {Zhang}, \citenamefont {Maffre},
  \citenamefont {Nienhaus},\ and\ \citenamefont {Parak}}]{Pino2014}%
  \BibitemOpen
  \bibfield  {author} {\bibinfo {author} {\bibfnamefont {P.}~\bibnamefont {del
  Pino}}, \bibinfo {author} {\bibfnamefont {B.}~\bibnamefont {Pelaz}}, \bibinfo
  {author} {\bibfnamefont {Q.}~\bibnamefont {Zhang}}, \bibinfo {author}
  {\bibfnamefont {P.}~\bibnamefont {Maffre}}, \bibinfo {author} {\bibfnamefont
  {G.~U.}\ \bibnamefont {Nienhaus}}, \ and\ \bibinfo {author} {\bibfnamefont
  {W.~J.}\ \bibnamefont {Parak}},\ }\href {\doibase 10.1039/C3MH00106G}
  {\bibfield  {journal} {\bibinfo  {journal} {Mater. Horiz.}\ }\textbf
  {\bibinfo {volume} {1}},\ \bibinfo {pages} {301} (\bibinfo {year}
  {2014})}\BibitemShut {NoStop}%
\bibitem [{\citenamefont {Brancolini}\ \emph {et~al.}(2012)\citenamefont
  {Brancolini}, \citenamefont {Kokh}, \citenamefont {Calzolai}, \citenamefont
  {Wade},\ and\ \citenamefont {Corni}}]{BKCW2012}%
  \BibitemOpen
  \bibfield  {author} {\bibinfo {author} {\bibfnamefont {G.}~\bibnamefont
  {Brancolini}}, \bibinfo {author} {\bibfnamefont {D.~B.}\ \bibnamefont
  {Kokh}}, \bibinfo {author} {\bibfnamefont {L.}~\bibnamefont {Calzolai}},
  \bibinfo {author} {\bibfnamefont {R.}~\bibnamefont {Wade}}, \ and\ \bibinfo
  {author} {\bibfnamefont {S.}~\bibnamefont {Corni}},\ }\href {\doibase
  10.1021/nn303444b} {\bibfield  {journal} {\bibinfo  {journal} {ACS Nano}\
  }\textbf {\bibinfo {volume} {6}},\ \bibinfo {pages} {9863} (\bibinfo {year}
  {2012})}\BibitemShut {NoStop}%
\bibitem [{\citenamefont {Ding}\ \emph {et~al.}(2013)\citenamefont {Ding},
  \citenamefont {Radic}, \citenamefont {Chen}, \citenamefont {Chen},
  \citenamefont {Geitner}, \citenamefont {Brown},\ and\ \citenamefont
  {Ke}}]{DRCC2013}%
  \BibitemOpen
  \bibfield  {author} {\bibinfo {author} {\bibfnamefont {F.}~\bibnamefont
  {Ding}}, \bibinfo {author} {\bibfnamefont {S.}~\bibnamefont {Radic}},
  \bibinfo {author} {\bibfnamefont {R.}~\bibnamefont {Chen}}, \bibinfo {author}
  {\bibfnamefont {P.}~\bibnamefont {Chen}}, \bibinfo {author} {\bibfnamefont
  {N.}~\bibnamefont {Geitner}}, \bibinfo {author} {\bibfnamefont
  {J.}~\bibnamefont {Brown}}, \ and\ \bibinfo {author} {\bibfnamefont
  {P.}~\bibnamefont {Ke}},\ }\href {\doibase 10.1039/c3nr02147e} {\bibfield
  {journal} {\bibinfo  {journal} {Nanoscale}\ }\textbf {\bibinfo {volume}
  {5}},\ \bibinfo {pages} {9162} (\bibinfo {year} {2013})}\BibitemShut
  {NoStop}%
\bibitem [{\citenamefont {Khan}, \citenamefont {Gupta},\ and\ \citenamefont
  {Nandi}(2013)}]{KGN2013}%
  \BibitemOpen
  \bibfield  {author} {\bibinfo {author} {\bibfnamefont {S.}~\bibnamefont
  {Khan}}, \bibinfo {author} {\bibfnamefont {A.}~\bibnamefont {Gupta}}, \ and\
  \bibinfo {author} {\bibfnamefont {C.}~\bibnamefont {Nandi}},\ }\href
  {\doibase 10.1021/jz401874u} {\bibfield  {journal} {\bibinfo  {journal} {J.
  Phys. Chem. Lett.}\ }\textbf {\bibinfo {volume} {4}},\ \bibinfo {pages}
  {3747} (\bibinfo {year} {2013})}\BibitemShut {NoStop}%
\bibitem [{\citenamefont {Tavanti}, \citenamefont {Pedone},\ and\ \citenamefont
  {Menziani}(2015)}]{Tavanti2015}%
  \BibitemOpen
  \bibfield  {author} {\bibinfo {author} {\bibfnamefont {F.}~\bibnamefont
  {Tavanti}}, \bibinfo {author} {\bibfnamefont {A.}~\bibnamefont {Pedone}}, \
  and\ \bibinfo {author} {\bibfnamefont {M.~C.}\ \bibnamefont {Menziani}},\
  }\href {\doibase 10.1039/c4nj01752h} {\bibfield  {journal} {\bibinfo
  {journal} {New J. Chem.}\ }\textbf {\bibinfo {volume} {39}},\ \bibinfo
  {pages} {2474} (\bibinfo {year} {2015})}\BibitemShut {NoStop}%
\bibitem [{\citenamefont {Vilaseca}, \citenamefont {Dawson},\ and\
  \citenamefont {Franzese}(2013)}]{VDF2013}%
  \BibitemOpen
  \bibfield  {author} {\bibinfo {author} {\bibfnamefont {P.}~\bibnamefont
  {Vilaseca}}, \bibinfo {author} {\bibfnamefont {K.}~\bibnamefont {Dawson}}, \
  and\ \bibinfo {author} {\bibfnamefont {G.}~\bibnamefont {Franzese}},\ }\href
  {\doibase 10.1039/C3SM50220A} {\bibfield  {journal} {\bibinfo  {journal}
  {Soft Matter}\ }\textbf {\bibinfo {volume} {9}},\ \bibinfo {pages} {6978}
  (\bibinfo {year} {2013})}\BibitemShut {NoStop}%
\bibitem [{\citenamefont {Bellion}\ \emph {et~al.}(2008)\citenamefont
  {Bellion}, \citenamefont {Santen}, \citenamefont {Mantz}, \citenamefont
  {Hoehl}, \citenamefont {Quinn}, \citenamefont {Nagel}, \citenamefont {Gilow},
  \citenamefont {Weitenberg}, \citenamefont {Schmitt},\ and\ \citenamefont
  {Jacobs}}]{Bellion2008}%
  \BibitemOpen
  \bibfield  {author} {\bibinfo {author} {\bibfnamefont {M.}~\bibnamefont
  {Bellion}}, \bibinfo {author} {\bibfnamefont {L.}~\bibnamefont {Santen}},
  \bibinfo {author} {\bibfnamefont {H.}~\bibnamefont {Mantz}}, \bibinfo
  {author} {\bibfnamefont {H.}~\bibnamefont {Hoehl}}, \bibinfo {author}
  {\bibfnamefont {A.}~\bibnamefont {Quinn}}, \bibinfo {author} {\bibfnamefont
  {A.}~\bibnamefont {Nagel}}, \bibinfo {author} {\bibfnamefont
  {C.}~\bibnamefont {Gilow}}, \bibinfo {author} {\bibfnamefont
  {C.}~\bibnamefont {Weitenberg}}, \bibinfo {author} {\bibfnamefont
  {Y.}~\bibnamefont {Schmitt}}, \ and\ \bibinfo {author} {\bibfnamefont
  {K.}~\bibnamefont {Jacobs}},\ }\href
  {http://stacks.iop.org/0953-8984/20/i=40/a=404226} {\bibfield  {journal}
  {\bibinfo  {journal} {J. Phys. Condens. Matter}\ }\textbf {\bibinfo {volume}
  {20}},\ \bibinfo {pages} {404226} (\bibinfo {year} {2008})}\BibitemShut
  {NoStop}%
\bibitem [{\citenamefont {Oberle}\ \emph {et~al.}(2015)\citenamefont {Oberle},
  \citenamefont {Yigit}, \citenamefont {Angioletti-Uberti}, \citenamefont
  {Dzubiella},\ and\ \citenamefont {Ballauff}}]{Oberle2015}%
  \BibitemOpen
  \bibfield  {author} {\bibinfo {author} {\bibfnamefont {M.}~\bibnamefont
  {Oberle}}, \bibinfo {author} {\bibfnamefont {C.}~\bibnamefont {Yigit}},
  \bibinfo {author} {\bibfnamefont {S.}~\bibnamefont {Angioletti-Uberti}},
  \bibinfo {author} {\bibfnamefont {J.}~\bibnamefont {Dzubiella}}, \ and\
  \bibinfo {author} {\bibfnamefont {M.}~\bibnamefont {Ballauff}},\ }\href
  {\doibase 10.1021/jp5119986} {\bibfield  {journal} {\bibinfo  {journal} {J.
  Phys. Chem. B}\ }\textbf {\bibinfo {volume} {119}},\ \bibinfo {pages} {3250 }
  (\bibinfo {year} {2015})}\BibitemShut {NoStop}%
\bibitem [{\citenamefont {Rabe}, \citenamefont {Verdes},\ and\ \citenamefont
  {Seeger}(2011)}]{Rabeetal2011}%
  \BibitemOpen
  \bibfield  {author} {\bibinfo {author} {\bibfnamefont {M.}~\bibnamefont
  {Rabe}}, \bibinfo {author} {\bibfnamefont {D.}~\bibnamefont {Verdes}}, \ and\
  \bibinfo {author} {\bibfnamefont {S.}~\bibnamefont {Seeger}},\ }\href
  {\doibase 10.1016/j.cis.2010.12.007} {\bibfield  {journal} {\bibinfo
  {journal} {Adv. Colloid Interface Sci.}\ }\textbf {\bibinfo {volume} {162}},\
  \bibinfo {pages} {87} (\bibinfo {year} {2011})}\BibitemShut {NoStop}%
\bibitem [{\citenamefont {Tozzini}(2005)}]{Toz2005}%
  \BibitemOpen
  \bibfield  {author} {\bibinfo {author} {\bibfnamefont {V.}~\bibnamefont
  {Tozzini}},\ }\href {\doibase 10.1016/j.sbi.2005.02.005} {\bibfield
  {journal} {\bibinfo  {journal} {Curr. Opin. Struct. Biol.}\ }\textbf
  {\bibinfo {volume} {15}},\ \bibinfo {pages} {144} (\bibinfo {year}
  {2005})}\BibitemShut {NoStop}%
\bibitem [{\citenamefont {Takada}(2012)}]{Tak2012}%
  \BibitemOpen
  \bibfield  {author} {\bibinfo {author} {\bibfnamefont {S.}~\bibnamefont
  {Takada}},\ }\href {\doibase 10.1016/j.sbi.2012.01.010} {\bibfield  {journal}
  {\bibinfo  {journal} {Curr. Opin. Struct. Biol.}\ }\textbf {\bibinfo {volume}
  {22}},\ \bibinfo {pages} {130} (\bibinfo {year} {2012})}\BibitemShut
  {NoStop}%
\bibitem [{\citenamefont {Noid}(2013)}]{Noid2013}%
  \BibitemOpen
  \bibfield  {author} {\bibinfo {author} {\bibfnamefont {W.~G.}\ \bibnamefont
  {Noid}},\ }\href {\doibase http://dx.doi.org/10.1063/1.4818908} {\bibfield
  {journal} {\bibinfo  {journal} {J. Chem. Phys.}\ }\textbf {\bibinfo {volume}
  {139}},\ \bibinfo {eid} {090901} (\bibinfo {year} {2013})}\BibitemShut
  {NoStop}%
\bibitem [{\citenamefont {{Hamaker}}(1937)}]{Hamaker1937}%
  \BibitemOpen
  \bibfield  {author} {\bibinfo {author} {\bibfnamefont {H.~C.}\ \bibnamefont
  {{Hamaker}}},\ }\href {\doibase 10.1016/S0031-8914(37)80203-7} {\bibfield
  {journal} {\bibinfo  {journal} {Physica}\ }\textbf {\bibinfo {volume} {4}},\
  \bibinfo {pages} {1058} (\bibinfo {year} {1937})}\BibitemShut {NoStop}%
\bibitem [{\citenamefont {Bereau}\ and\ \citenamefont
  {Deserno}(2009)}]{BerDes2009}%
  \BibitemOpen
  \bibfield  {author} {\bibinfo {author} {\bibfnamefont {T.}~\bibnamefont
  {Bereau}}\ and\ \bibinfo {author} {\bibfnamefont {M.}~\bibnamefont
  {Deserno}},\ }\href {\doibase 10.1063/1.3152842} {\bibfield  {journal}
  {\bibinfo  {journal} {J. Chem. Phys.}\ }\textbf {\bibinfo {volume} {130}},\
  \bibinfo {pages} {235106} (\bibinfo {year} {2009})}\BibitemShut {NoStop}%
\bibitem [{\citenamefont {Miyazawa}\ and\ \citenamefont
  {Jernigan}(1996)}]{MiyJer1996}%
  \BibitemOpen
  \bibfield  {author} {\bibinfo {author} {\bibfnamefont {S.}~\bibnamefont
  {Miyazawa}}\ and\ \bibinfo {author} {\bibfnamefont {R.}~\bibnamefont
  {Jernigan}},\ }\href {\doibase 10.1006/jmbi.1996.0114} {\bibfield  {journal}
  {\bibinfo  {journal} {J. Mol. Biol.}\ }\textbf {\bibinfo {volume} {256}},\
  \bibinfo {pages} {623} (\bibinfo {year} {1996})}\BibitemShut {NoStop}%
\bibitem [{\citenamefont {Kim}\ \emph {et~al.}(2008)\citenamefont {Kim},
  \citenamefont {Tang}, \citenamefont {Clore},\ and\ \citenamefont
  {Hummer}}]{KTMH2008}%
  \BibitemOpen
  \bibfield  {author} {\bibinfo {author} {\bibfnamefont {Y.}~\bibnamefont
  {Kim}}, \bibinfo {author} {\bibfnamefont {C.}~\bibnamefont {Tang}}, \bibinfo
  {author} {\bibfnamefont {G.}~\bibnamefont {Clore}}, \ and\ \bibinfo {author}
  {\bibfnamefont {G.}~\bibnamefont {Hummer}},\ }\href {\doibase
  10.1073/pnas.0802460105} {\bibfield  {journal} {\bibinfo  {journal} {Proc.
  Natl. Acad. Sci. USA}\ }\textbf {\bibinfo {volume} {105}},\ \bibinfo {pages}
  {12855} (\bibinfo {year} {2008})}\BibitemShut {NoStop}%
\bibitem [{\citenamefont {Kim}\ and\ \citenamefont
  {Hummer}(2008)}]{KimHum2008}%
  \BibitemOpen
  \bibfield  {author} {\bibinfo {author} {\bibfnamefont {Y.}~\bibnamefont
  {Kim}}\ and\ \bibinfo {author} {\bibfnamefont {G.}~\bibnamefont {Hummer}},\
  }\href {\doibase 10.1016/j.jmb.2007.11.063} {\bibfield  {journal} {\bibinfo
  {journal} {J. Mol. Biol.}\ }\textbf {\bibinfo {volume} {375}},\ \bibinfo
  {pages} {1416} (\bibinfo {year} {2008})}\BibitemShut {NoStop}%
\bibitem [{\citenamefont {Wei}\ and\ \citenamefont
  {Knotts}(2013)}]{WeiKno2013}%
  \BibitemOpen
  \bibfield  {author} {\bibinfo {author} {\bibfnamefont {S.}~\bibnamefont
  {Wei}}\ and\ \bibinfo {author} {\bibfnamefont {T.}~\bibnamefont {Knotts}},\
  }\href {\doibase 10.1063/1.4819131} {\bibfield  {journal} {\bibinfo
  {journal} {J. Chem. Phys.}\ }\textbf {\bibinfo {volume} {139}},\ \bibinfo
  {pages} {095102} (\bibinfo {year} {2013})}\BibitemShut {NoStop}%
\bibitem [{\citenamefont {Agashe}\ \emph {et~al.}(2005)\citenamefont {Agashe},
  \citenamefont {Raut}, \citenamefont {Stuart},\ and\ \citenamefont
  {Latour}}]{ARS2005}%
  \BibitemOpen
  \bibfield  {author} {\bibinfo {author} {\bibfnamefont {M.}~\bibnamefont
  {Agashe}}, \bibinfo {author} {\bibfnamefont {V.}~\bibnamefont {Raut}},
  \bibinfo {author} {\bibfnamefont {S.}~\bibnamefont {Stuart}}, \ and\ \bibinfo
  {author} {\bibfnamefont {R.}~\bibnamefont {Latour}},\ }\href@noop {}
  {\bibfield  {journal} {\bibinfo  {journal} {Langmuir}\ }\textbf {\bibinfo
  {volume} {21}},\ \bibinfo {pages} {1103} (\bibinfo {year}
  {2005})}\BibitemShut {NoStop}%
\bibitem [{\citenamefont {Sun}, \citenamefont {Welsh},\ and\ \citenamefont
  {Latour}(2005)}]{SWL2005}%
  \BibitemOpen
  \bibfield  {author} {\bibinfo {author} {\bibfnamefont {Y.}~\bibnamefont
  {Sun}}, \bibinfo {author} {\bibfnamefont {W.}~\bibnamefont {Welsh}}, \ and\
  \bibinfo {author} {\bibfnamefont {R.}~\bibnamefont {Latour}},\ }\href
  {\doibase 10.1021/la046932o} {\bibfield  {journal} {\bibinfo  {journal}
  {Langmuir}\ }\textbf {\bibinfo {volume} {21}},\ \bibinfo {pages} {5616}
  (\bibinfo {year} {2005})}\BibitemShut {NoStop}%
\bibitem [{\citenamefont {Kokh}\ \emph {et~al.}(2010)\citenamefont {Kokh},
  \citenamefont {Corni}, \citenamefont {Winn}, \citenamefont {Hoefling},
  \citenamefont {Gottschalk},\ and\ \citenamefont {Wade}}]{Dariaetal2010}%
  \BibitemOpen
  \bibfield  {author} {\bibinfo {author} {\bibfnamefont {D.}~\bibnamefont
  {Kokh}}, \bibinfo {author} {\bibfnamefont {S.}~\bibnamefont {Corni}},
  \bibinfo {author} {\bibfnamefont {P.}~\bibnamefont {Winn}}, \bibinfo {author}
  {\bibfnamefont {M.}~\bibnamefont {Hoefling}}, \bibinfo {author}
  {\bibfnamefont {K.}~\bibnamefont {Gottschalk}}, \ and\ \bibinfo {author}
  {\bibfnamefont {R.}~\bibnamefont {Wade}},\ }\href {\doibase
  10.1021/ct100086j} {\bibfield  {journal} {\bibinfo  {journal} {J. Chem.
  Theory Comput.}\ }\textbf {\bibinfo {volume} {6}},\ \bibinfo {pages} {1753}
  (\bibinfo {year} {2010})}\BibitemShut {NoStop}%
\bibitem [{\citenamefont {Lesniak}\ \emph {et~al.}(2010)\citenamefont
  {Lesniak}, \citenamefont {Campbell}, \citenamefont {Monopoli}, \citenamefont
  {Lynch}, \citenamefont {Salvati},\ and\ \citenamefont {Dawson}}]{1-36}%
  \BibitemOpen
  \bibfield  {author} {\bibinfo {author} {\bibfnamefont {A.}~\bibnamefont
  {Lesniak}}, \bibinfo {author} {\bibfnamefont {A.}~\bibnamefont {Campbell}},
  \bibinfo {author} {\bibfnamefont {M.~P.}\ \bibnamefont {Monopoli}}, \bibinfo
  {author} {\bibfnamefont {I.}~\bibnamefont {Lynch}}, \bibinfo {author}
  {\bibfnamefont {A.}~\bibnamefont {Salvati}}, \ and\ \bibinfo {author}
  {\bibfnamefont {K.~A.}\ \bibnamefont {Dawson}},\ }\href@noop {} {\bibfield
  {journal} {\bibinfo  {journal} {Biomaterials}\ }\textbf {\bibinfo {volume}
  {31}},\ \bibinfo {pages} {9511} (\bibinfo {year} {2010})}\BibitemShut
  {NoStop}%
\bibitem [{\citenamefont {Limbach}\ \emph {et~al.}(2006)\citenamefont
  {Limbach}, \citenamefont {Arnold}, \citenamefont {Mann},\ and\ \citenamefont
  {Holm}}]{Espresso}%
  \BibitemOpen
  \bibfield  {author} {\bibinfo {author} {\bibfnamefont {H.}~\bibnamefont
  {Limbach}}, \bibinfo {author} {\bibfnamefont {A.}~\bibnamefont {Arnold}},
  \bibinfo {author} {\bibfnamefont {B.}~\bibnamefont {Mann}}, \ and\ \bibinfo
  {author} {\bibfnamefont {C.}~\bibnamefont {Holm}},\ }\href {\doibase
  10.1016/j.cpc.2005.10.005} {\bibfield  {journal} {\bibinfo  {journal}
  {Comput. Phys. Commun.}\ }\textbf {\bibinfo {volume} {174}},\ \bibinfo
  {pages} {704} (\bibinfo {year} {2006})}\BibitemShut {NoStop}%
\bibitem [{\citenamefont {Chen}\ \emph {et~al.}(2003)\citenamefont {Chen},
  \citenamefont {Huang}, \citenamefont {Lin}, \citenamefont {Lin},\ and\
  \citenamefont {Chan}}]{CHL2003}%
  \BibitemOpen
  \bibfield  {author} {\bibinfo {author} {\bibfnamefont {W.}~\bibnamefont
  {Chen}}, \bibinfo {author} {\bibfnamefont {H.}~\bibnamefont {Huang}},
  \bibinfo {author} {\bibfnamefont {C.}~\bibnamefont {Lin}}, \bibinfo {author}
  {\bibfnamefont {F.}~\bibnamefont {Lin}}, \ and\ \bibinfo {author}
  {\bibfnamefont {Y.}~\bibnamefont {Chan}},\ }\href {\doibase
  10.1021/la034783o} {\bibfield  {journal} {\bibinfo  {journal} {Langmuir}\
  }\textbf {\bibinfo {volume} {19}},\ \bibinfo {pages} {9395} (\bibinfo {year}
  {2003})}\BibitemShut {NoStop}%
\bibitem [{\citenamefont {Lacerda}\ \emph {et~al.}(2010)\citenamefont
  {Lacerda}, \citenamefont {Park}, \citenamefont {Meuse}, \citenamefont
  {Pristinski}, \citenamefont {Becker}, \citenamefont {Karim},\ and\
  \citenamefont {Douglas}}]{LPJ2010}%
  \BibitemOpen
  \bibfield  {author} {\bibinfo {author} {\bibfnamefont {S.}~\bibnamefont
  {Lacerda}}, \bibinfo {author} {\bibfnamefont {J.}~\bibnamefont {Park}},
  \bibinfo {author} {\bibfnamefont {C.}~\bibnamefont {Meuse}}, \bibinfo
  {author} {\bibfnamefont {D.}~\bibnamefont {Pristinski}}, \bibinfo {author}
  {\bibfnamefont {M.}~\bibnamefont {Becker}}, \bibinfo {author} {\bibfnamefont
  {A.}~\bibnamefont {Karim}}, \ and\ \bibinfo {author} {\bibfnamefont
  {J.}~\bibnamefont {Douglas}},\ }\href {\doibase 10.1021/nn9011187} {\bibfield
   {journal} {\bibinfo  {journal} {ACS Nano}\ }\textbf {\bibinfo {volume}
  {4}},\ \bibinfo {pages} {365} (\bibinfo {year} {2010})}\BibitemShut {NoStop}%
\bibitem [{\citenamefont {Vroman}(1962)}]{Vroman1962}%
  \BibitemOpen
  \bibfield  {author} {\bibinfo {author} {\bibfnamefont {L.}~\bibnamefont
  {Vroman}},\ }\href {\doibase 10.1038/196476a0} {\bibfield  {journal}
  {\bibinfo  {journal} {Nature}\ }\textbf {\bibinfo {volume} {196}},\ \bibinfo
  {pages} {476} (\bibinfo {year} {1962})}\BibitemShut {NoStop}%
\bibitem [{\citenamefont {LeDuc}, \citenamefont {Vroman},\ and\ \citenamefont
  {Leonard}(1995)}]{DeLuc1995}%
  \BibitemOpen
  \bibfield  {author} {\bibinfo {author} {\bibfnamefont {C.~A.}\ \bibnamefont
  {LeDuc}}, \bibinfo {author} {\bibfnamefont {L.}~\bibnamefont {Vroman}}, \
  and\ \bibinfo {author} {\bibfnamefont {E.~F.}\ \bibnamefont {Leonard}},\
  }\href@noop {} {\bibfield  {journal} {\bibinfo  {journal} {Ind. Eng. Chem.
  Res.}\ }\textbf {\bibinfo {volume} {34}},\ \bibinfo {pages} {3488} (\bibinfo
  {year} {1995})}\BibitemShut {NoStop}%
\bibitem [{\citenamefont {Ortega-Vinuesa}\ and\ \citenamefont
  {Hidalgo-Alvarez}(1995)}]{Ortega1995}%
  \BibitemOpen
  \bibfield  {author} {\bibinfo {author} {\bibfnamefont {J.}~\bibnamefont
  {Ortega-Vinuesa}}\ and\ \bibinfo {author} {\bibfnamefont {R.}~\bibnamefont
  {Hidalgo-Alvarez}},\ }\href@noop {} {\bibfield  {journal} {\bibinfo
  {journal} {Biotechnol. Bioeng.}\ }\textbf {\bibinfo {volume} {47}},\ \bibinfo
  {pages} {633} (\bibinfo {year} {1995})}\BibitemShut {NoStop}%
\bibitem [{\citenamefont {Holmberg}\ and\ \citenamefont
  {Hou}(2009)}]{Holmberg2009}%
  \BibitemOpen
  \bibfield  {author} {\bibinfo {author} {\bibfnamefont {M.}~\bibnamefont
  {Holmberg}}\ and\ \bibinfo {author} {\bibfnamefont {X.}~\bibnamefont {Hou}},\
  }\href {\doibase 10.1021/la8031978} {\bibfield  {journal} {\bibinfo
  {journal} {Langmuir}\ }\textbf {\bibinfo {volume} {25}},\ \bibinfo {pages}
  {2081} (\bibinfo {year} {2009})}\BibitemShut {NoStop}%
\bibitem [{\citenamefont {Koehler}, \citenamefont {Schmid},\ and\ \citenamefont
  {Settanni}(2014)}]{Schmid2014}%
  \BibitemOpen
  \bibfield  {author} {\bibinfo {author} {\bibfnamefont {S.}~\bibnamefont
  {Koehler}}, \bibinfo {author} {\bibfnamefont {F.}~\bibnamefont {Schmid}}, \
  and\ \bibinfo {author} {\bibfnamefont {G.}~\bibnamefont {Settanni}},\
  }\href@noop {} {\bibfield  {journal} {\bibinfo  {journal} {NIC Series}\
  }\textbf {\bibinfo {volume} {47}},\ \bibinfo {pages} {117} (\bibinfo {year}
  {2014})}\BibitemShut {NoStop}%
\bibitem [{\citenamefont {Brandt}\ and\ \citenamefont
  {Lyubartsev}(2014)}]{Brandt2014}%
  \BibitemOpen
  \bibfield  {author} {\bibinfo {author} {\bibfnamefont {E.}~\bibnamefont
  {Brandt}}\ and\ \bibinfo {author} {\bibfnamefont {A.~P.}\ \bibnamefont
  {Lyubartsev}},\ }\href@noop {} {\bibfield  {journal} {\bibinfo  {journal}
  {Biophys. J.}\ }\textbf {\bibinfo {volume} {106}},\ \bibinfo {pages} {208a}
  (\bibinfo {year} {2014})}\BibitemShut {NoStop}%
\bibitem [{\citenamefont {Brandt}\ and\ \citenamefont
  {Lyubartsev}(2015)}]{Brandt2015}%
  \BibitemOpen
  \bibfield  {author} {\bibinfo {author} {\bibfnamefont {E.}~\bibnamefont
  {Brandt}}\ and\ \bibinfo {author} {\bibfnamefont {A.~P.}\ \bibnamefont
  {Lyubartsev}},\ }\href {\doibase 10.1021/acs.jpcc.5b02670} {\bibfield
  {journal} {\bibinfo  {journal} {J. Phys. Chem. C}\ } (\bibinfo {year}
  {2015}),\ 10.1021/acs.jpcc.5b02670}\BibitemShut {NoStop}%
\end{thebibliography}%

\end{document}